\documentclass[%
reprint,
superscriptaddress,
nofootinbib,
nobibnotes,
 amsmath,amssymb,
 aps,
 prd,
 longbibliography,
 twocolumn,
]{revtex4-2}

\usepackage{float}
\usepackage[percent]{overpic}
\usepackage{graphicx, comment}
\usepackage[caption=false]{subfig}
\usepackage{dcolumn}
\usepackage{bm}
\usepackage{xcolor}
\usepackage[mathlines]{lineno}
\usepackage[normalem]{ulem}
\usepackage{amsfonts}
\usepackage{url}
\usepackage{lineno}
\usepackage{url}
\usepackage{hyperref}

\begin{document}

\newlength{\figwidth}
\newlength{\fighalfwidth}
\setlength{\figwidth}{0.95\textwidth}
\setlength{\fighalfwidth}{0.5\textwidth}
\newcommand{\blue}[1]{{\textcolor{blue}{#1}}}
\newcommand{\red}[1]{{\textcolor{red}{#1}}}
\newcommand{\fixme}[1]{\textit{\textcolor{red}{Fixme: #1}}}

\newcommand{\Ar}{\mathrm{Ar}}


\title{Physics Prospects with MeV Neutrino-Argon Charged Current Interactions using Enhanced Photon Detection in Future LArTPCs}

\newcommand{\SBU}{Stony Brook University, SUNY, Stony Brook, NY, 11794, USA}
\newcommand{\BNL}{Brookhaven National Laboratory, Upton, NY, 11973, USA}
\newcommand{\FSU}{Florida State University, Tallahassee, FL 32306, USA}
\newcommand{\ITA}{Instituto Tecnologico de Aeronautica (ITA), São José dos Campos/SP, 12228, Brasil}
\newcommand{\FNAL}{Fermi National Accelerator Laboratory, Batavia, IL, 60510, U.S.A.}
\newcommand{\MADA}{University of Antananarivo, BP 566, Antananarivo-101, Madagascar}

\author{Wei Shi} 
\email{wei.shi.1@stonybrook.edu} 
\affiliation{\SBU}
\author{Xuyang Ning} \affiliation{\BNL}
\author{Daniel Pershey} 
\email{dpershey@fsu.edu} 
\affiliation{\FSU}
\author{Franciole Marinho} \affiliation{\ITA}
\author{A. Fleuri}\affiliation{\MADA}
\author{Ciro Riccio} \affiliation{\SBU}
\author{Jay Hyun Jo} \affiliation{\BNL}
\author{Chao Zhang} \affiliation{\BNL}
\author{Flavio Cavanna} \affiliation{\FNAL}

\date{\today}
\begin{abstract}
We investigate MeV-scale electron neutrino charged current interactions in a liquid argon time projection chamber equipped with an enhanced photon detection system. Using simulations of deposited energy in charge and light calorimetry, we explore the potential for dual calorimetric neutrino energy reconstruction. We found energy reconstruction based on light-only calorimetry has a better resolution than combined charge and light calorimetry when hadrons are produced in these events. Meanwhile, enhanced light detection offers improved nanosecond timing resolution and broad optical coverage, enabling neutron tagging and identification of delayed low-energy gamma emissions. These advancements open new avenues in low-energy neutrino physics in next-generation LArTPCs.
\end{abstract}

\maketitle


\section{Introduction}
\label{sec:intro}

Noble liquid time projection chambers (TPCs) are widely employed for detecting rare and low-energy signals, such as neutrinos and dark matter interactions. Liquid argon (LAr) is particularly favored for its scalability and excellent tracking and calorimetric capabilities~\cite{CalorimetryBibleFabjan}. Experiments such as LArIAT, ArgoNeuT, and MicroBooNE~\cite{ArgoNeuT-MeV, LArIAT_LowE_electron, MicroBooNE_MeV} have demonstrated potentials of reconstructing MeV-scale activities. This capability is crucial for the detection and identification of neutrinos with energies as low as a few MeV, where electron neutrino ($\nu_{e}$) charged current (CC) interactions on argon nuclei serve as the primary detection channel at DUNE~\cite{DUNE_FD_TDR_Physics}. These neutrinos originate from supernova bursts~\cite{Gil-Botella_2011, DUNE_SNB}, solar fusion reactions~\cite{Bahcall:1989ks, SolarMSW, FRIEDLAND2004347}, and the diffuse supernova neutrino background (DSNB)~\cite{Ciaran_NeutrinoFloor, SURF_atm_flux, SK_DSNB}.

Argon (Ar) and xenon (Xe) emit vacuum ultra-violet (VUV) scintillation light, peaking at 127 nm and 178 nm, respectively, when charged particles deposit energy inside~\cite{NobleliquidScintillationlight, LAr_light}. In noble liquid TPCs, charge signals from ionization can be collected in the presence of an electric field. The anti-correlation between charge and light signals, well-established in liquid xenon (LXe) experiments, has been used to improve energy reconstruction in searches for neutrinoless double beta decay and dark matter~\cite{EXO200, Xenon100}. In the XENON experiment, the hypothesized weakly interacting massive particle dark matter candidate would be detected through nuclear or electron recoils that deposit tens of keV energy~\cite{XenonNT_NuclearRecoil, XenonNT_ElectronRecoil}. EXO-200~\cite{EXO200} has performed measurements of the absolute yields of charge and light in LXe for electron recoils up to MeV. 

While charge detection in LArTPCs has matured~\cite{MicroBooNE:2016pwy, ICARUS-T600, ICARUS-FNAL, SBND:2020scp, DUNE:2021hwx}, optical coverage and light yield (LY) remain limited. In LArTPC neutrino experiments, event reconstruction primarily uses the charge signal while the light signal is mostly used to provide timing information for non-beam neutrino events. Matching charge and light signals to facilitate cosmic background rejection has been explored at MicroBooNE~\cite{MicroBooNE_QLMatching_WireCell}. It is anticipated an enhanced photon detection system in future LArTPCs can offer more capabilities for physics with MeV $\nu_{e}$-Ar CC interactions. The recently proposed Aluminum Profiles with Embedded X-Arapuca (APEX)~\cite{DUNE_Phase2} for DUNE's third and fourth far detector LArTPC modules suggests optical coverage of up to 60\% ($\sim$2000 m$^2$) and an average LY of $\sim$180 photoelectrons per MeV (PE/MeV), presenting a great potential for low-energy physics.

In this work, we study the prospects of MeV neutrino physics in a LArTPC with enhanced photon detection. Key improvements include increased MeV energy resolution from light calorimetry, excellent timing resolution, and extensive optical coverage. These advancements facilitate the identification of delayed energy deposits and particle tagging, improving signal detection and background rejection, particularly in searches for the DSNB as well as many other interesting physics topics~\cite{SolarMSW, DUNE_FD_TDR_Physics}. The insights gained in this study will guide the design and calibration of photon detection systems in future LArTPCs.

The paper is structured as follows: Sec~\ref{sec:sim} describes the simulation setup and energy smearing effects in LAr. Sec~\ref{sec:QLDet} details the simulated charge and light detection processes. The expected benefits of an enhanced photon detection system are discussed in Sec~\ref{sec:ereco}. An example application to DSNB searches is discussed in Sec~\ref{sec:app}. A final summary is provided in Sec~\ref{sec:summary}.

\section{Simulation}
\label{sec:sim}

The MeV-scale $\nu_{e}$ CC interactions with an Ar nucleus are simulated using the MARLEY v1.2.0 event generator~\cite{Marley_main, Marley_nueArCC}. Monoenergetic events are generated at energies ranging  from 5 MeV to 50 MeV in 5 MeV increments. These events originate at the coordinate origin and propagate in the $\hat{z}$ direction. For each energy, 1000 events are generated. The final-state particles from the neutrino interaction are tracked using Geant4 v4.10.6.p01~\cite{Geant4_1, Geant4_2, Geant4_3}, interfaced via the edep-sim package~\cite{edep-sim}. The default $\tt{QGSP\_BERT}$ physics list is used. The neutron killer is deactivated to properly simulate neutron and its captures. The simulation is conducted in a large LAr volume, measuring 200 meters in each dimension, to ensure full containment of all simulated events. The deposited energy is recorded with a maximum step size of 0.5 cm, along with the corresponding timing information for each energy deposit. All electrons from energy deposition into charge calorimetry are assumed to drift under an infinite lifetime and be collected by charge readout.

\begin{figure}[htp]
  \centering
    \includegraphics[width=0.9\columnwidth]{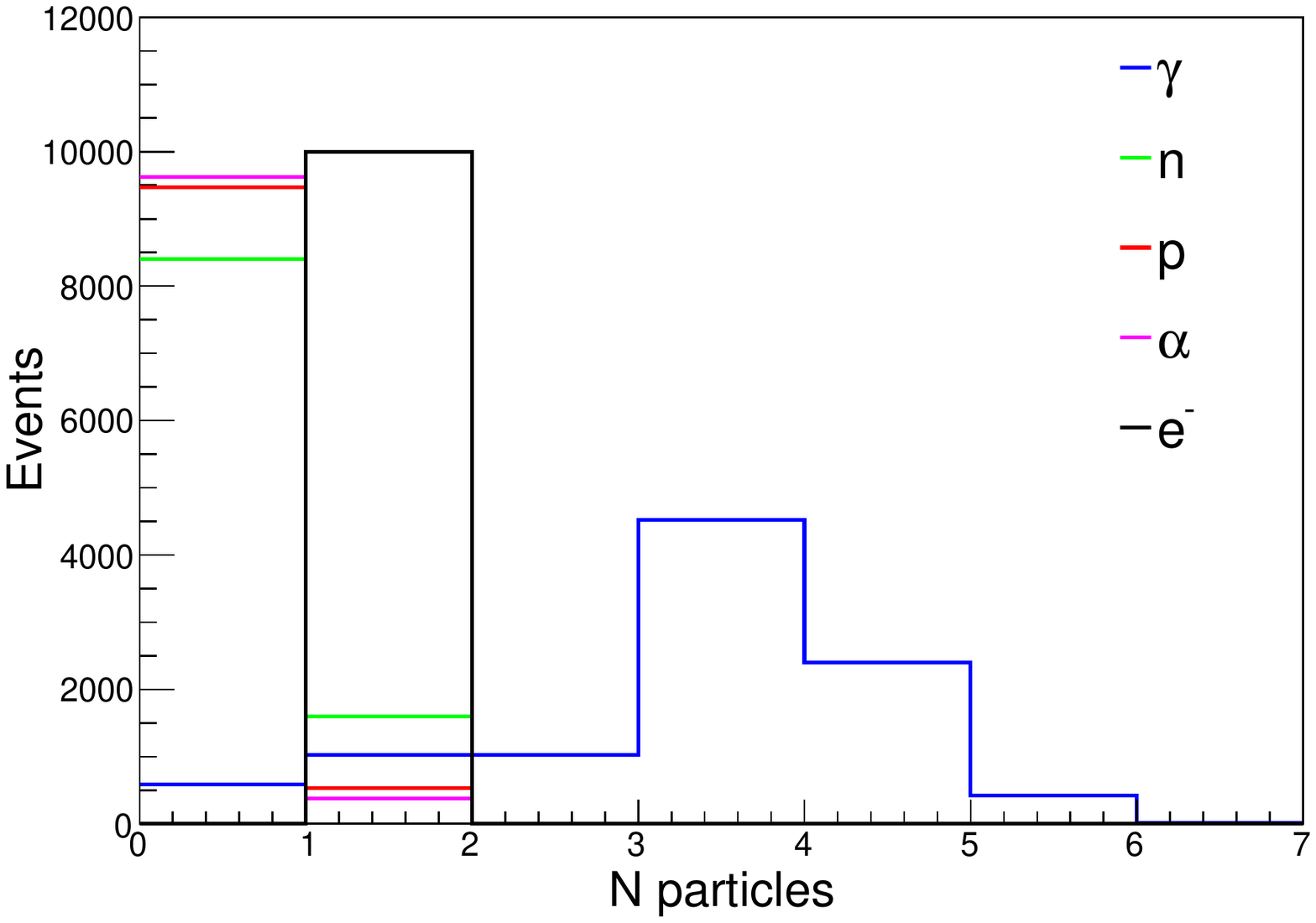}
  \caption{Distribution of multiplicity for final state particles from $\nu_{e}$-Ar CC interactions for incoming $\nu_{e}$ with energy in 5 - 50 MeV following a uniform distribution with mono-energetic spacing.}
  \label{fig:GenNp}
\end{figure}

The multiplicity distribution of final-state particles for the MeV $\nu_{e}$-Ar CC events is shown in Fig.~\ref{fig:GenNp}. The electron carries most of the energy from the incoming $\nu_{e}$ and primarily loses energy through scattering, ionization, and bremsstrahlung in LAr, depending on the initial neutrino energy. The resulting electron track in the LArTPC ranges in length from sub-centimeter to tens of centimeters. Gamma rays ($\gamma$s) from the de-excitation of the final-state excited potassium-40 nucleus ($^{40}$K$^\ast$) have energies below 5 MeV. These $\gamma$s primarily lose energy through Compton scattering, pair production, and photoelectric interactions. Their energy deposits appear as blip-like objects, typically spanning a spherical region with a diameter of approximately 1 meter in the LArTPC~\cite{MeVNueArCC}. At higher $\nu_{e}$ energy, one or more hadrons---primarily neutrons, protons, and $\alpha$ particles---can be knocked out of the Ar nucleus. 

To study energy smearing, we define the available energy, $E_{\text{avail}}$, for outgoing final-state particles from the primary $\nu_{e}$ CC interaction modeled by MARLEY. For electrons and $\gamma$s, $E_{\text{avail}}$ is defined as the particle's total energy in the simulation. For hadrons or nuclei, $E_{\text{avail}}$ corresponds to their kinetic energy. The total $E_{\text{avail}}$ of a $\nu_{e}$~CC event is then the sum of $E_{\text{avail}}$ from all final-state particles. 

\begin{figure}[htp]
  \centering  \includegraphics[width=0.9\columnwidth]{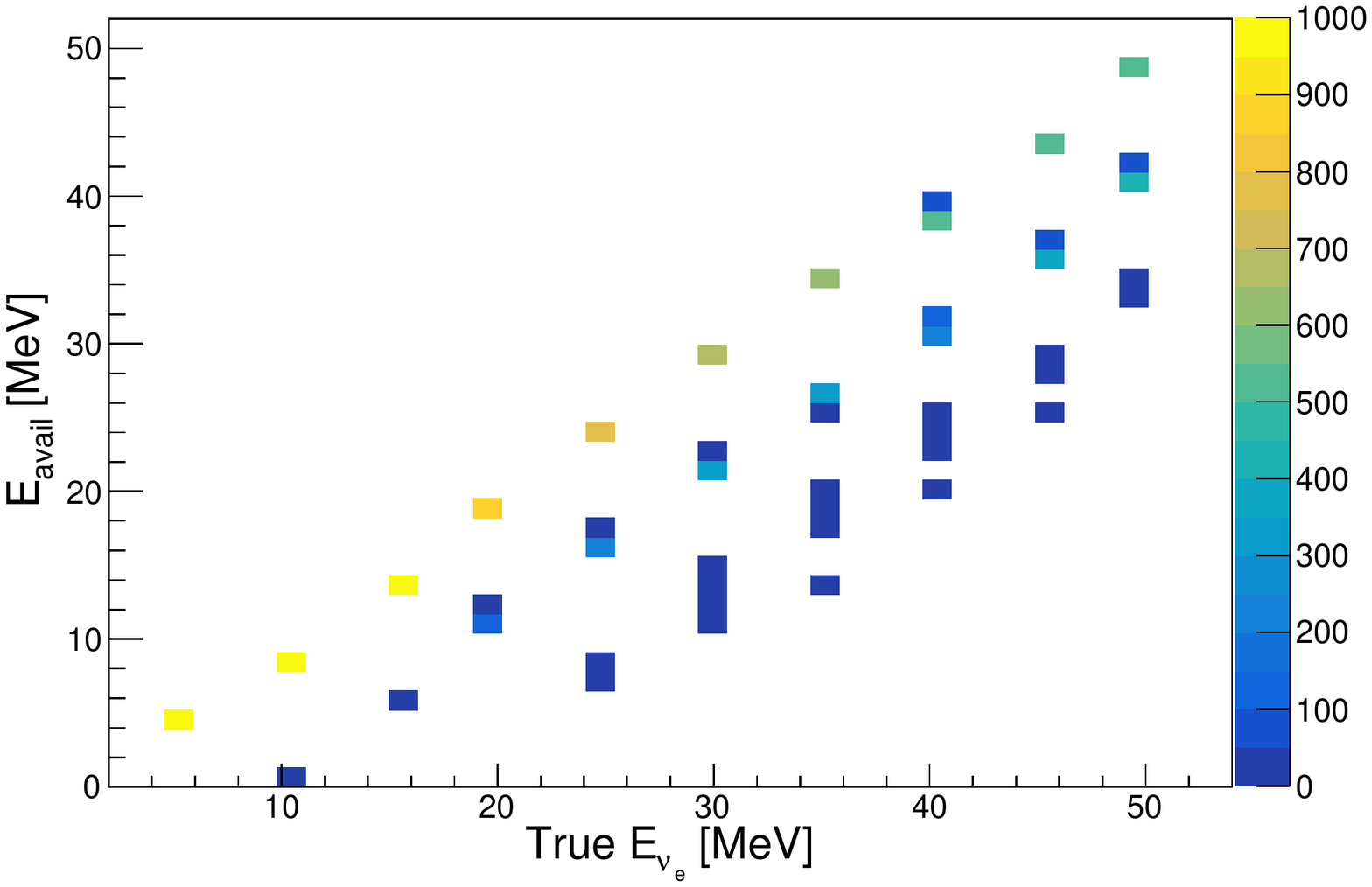}
  \centering  \includegraphics[width=0.9\columnwidth]{hadronfraction.pdf}
  \caption{Top: Distribution of $E_{\text{avail}}$ and true $E_{\nu_{e}}$ for generated $\nu_{e}$-Ar CC interactions for incoming $\nu_{e}$ with energy in 5 - 50 MeV. Bottom: Fraction of hadron production events as a function of true neutrino energy. Most hadron emission events produce only a neutron (green, $\text{N}_{\text{n}}$ = 1), a proton (red, $\text{N}_{\text{p}}$ = 1), or an alpha particle (pink, $\text{N}_{\alpha}$ = 1). The black line shows the fraction of events with a production of any of the three hadrons.} 
  \label{fig:2DEavailEtrue}
\end{figure}

The two-dimensional distribution of the true $\nu_{e}$ energy ($E_{\nu_{e}}$) versus $E_{\text{avail}}$ is shown in the top plot of Fig.~\ref{fig:2DEavailEtrue}. Starting at 10 MeV, an energy loss of approximately 8 MeV due to nuclear binding energy is observed when hadrons are ejected from the Ar nucleus. This results in a secondary peak in the $E_{\text{avail}}$ distribution, offset by 8 MeV from the primary peak (cf. Fig.~\ref{fig:DetectedQL} bottom plot). The fraction of hadron production events as a function of true $E_{\nu_{e}}$ is shown in the bottom plot of Fig.~\ref{fig:2DEavailEtrue}. All hadrons considered in the study below--neutrons, protons, and $\alpha$ particles--have higher production rates as neutrino energy increases. Overall, hadron knockout events constitute about 25\% of all $\nu_{e}$~CC interactions across the 5-50 MeV range, with neutrons being the most common ($\sim$65\%). The neutron’s 7.9 MeV binding energy escapes detection. Protons and $\alpha$ particles also appear in the final-state but are less common, requiring 7.6 MeV and 7.1 MeV, respectively, to overcome their binding energies.  Additionally, a systematic energy shift of approximately $-$1 MeV is observed in the MARLEY event generator, accounting for the binding energy change from the initial Ar nucleus to the K nucleus~\cite{Marley_nueArCC}. Nuclear quenching effects on the small, keV-scale recoil energy of the daughter $^{40}$K nucleus are negligible and are therefore ignored in this study. 

For each final-state particle exiting the nucleus and interacting with the bulk argon, we define its deposited energy, $E_{\text{depo}}$, which includes
energy deposition from all daughter particles involved in scattering, decay, de-excitation, etc. The primary energy smearing effect at this stage arises from neutron capture on the argon-40 ($^{40}$Ar) nucleus, where an additional 6.1 MeV is deposited due to the de-excitation of the resulting argon-41 ($^{41}$Ar) nucleus~\cite{Ar41}. In this study, only neutron captures on $^{40}$Ar are simulated, and all neutrons in the simulated events are eventually captured. The $^{41}$Ar de-excitation releases several low energy gammas which may be separated out using timing and spatial information but is not considered here.

The secondary peak in the $E_{\text{avail}}$ distribution (Fig.~\ref{fig:DetectedQL} bottom plot) further splits into two smaller secondary peaks in the $E_{\text{depo}}$ distribution. Events containing neutrons are clustered approximately 1.8 MeV below the primary $E_{\text{depo}}$ peak due to the combined effects of undetected neutron binding energy and the additional energy released from neutron capture. Events containing hadrons other than neutron cluster at $E_{\text{depo}}$ regions similar to the secondary peak in the $E_{\text{avail}}$ distribution. Furthermore, the $E_{\text{depo}}$ of most events is shifted downward by 0.51 MeV relative to $E_{\text{avail}}$, corresponding to the rest mass of the final-state electron.

In summary, two energy smearing mechanisms are relevant for MeV-scale $\nu_{e}$ CC events. The first is the hadron knockout effect from the primary $\nu_{e}$-Ar interaction, where a binding energy of approximately 8 MeV remains undetected in higher energy $\nu_{e}$ events. The second arises from the neutron capture, which introduces an additional 6.1 MeV deposition due to the $\gamma$ cascade from the de-excitation of the $^{41}$Ar. These two mechanisms create secondary peaks at lower $E_{\text{depo}}$ (Fig.~\ref{fig:DetectedQL} bottom plot), broadening the energy distribution and impacting low-energy physics searches.

\section{Charge and Light Detection}
\label{sec:QLDet}

The mechanisms for energy dissipation as charge and light in LAr have been extensively studied in previous literature~\cite {Selfcompensatinglight4GeV, LAr_Charge,  LAr_light, LArQL}. In this study, the energy deposited as charge and light is simulated using the Birks model~\cite{Birkslaw}, which describes the recombination of ionized electrons with noble element ions, leading to the release of additional scintillation light. For each simulated energy deposit, the corresponding linear energy transfer, $dE/dx$, is converted to the linear charge transfer, $dQ/dx$, using the following equation:
\begin{equation}\label{eq:dQ}
\frac{{dQ}}{{dx}} = \frac{{dE}}{{dx}} \times 0.83 \times {R}_{\text{c}}
\end{equation}
Here, 0.83 is the measured fraction of deposited energy that initially undergoes ionization~\cite{IonizationfracLAr}, while the recombination factor, ${R}_{\text{c}}$, is derived from the Birks model: 
\begin{equation}\label{eq:birks}
{R}_{\text{c}} = \frac{A}{1 + k/\epsilon \times {dE/dx}}
\end{equation}
where $A =$ 0.8, $k$ is 0.0486 $\frac{\text{g}}{\text{MeV}\ \text{cm}^2} \frac{\text{kV}}{\text{cm}}$~\cite{ICARUSrecombination, Recombination}. The drift electric field, $\epsilon =$ 0.5 kV/cm, is used in this simulation. 

Two charge detection thresholds, 75 keV and 500 keV, are considered and applied to the simulated charge calorimetry. The 75 keV threshold represents an optimistic detection limit, similarly used in Ref \cite{MeVNueArCC}. MicroBooNE has achieved a 210 keV charge detection threshold, where the reconstruction efficiency for low energy electrons reaches 50\% of its maximum achievable value in a “low-threshold” configuration on its charge collection plane~\cite{PhysRevD.109.052007}. The 500 keV threshold serves as a more conservative benchmark. For each single energy deposit ($dE$) into charge calorimetry ($dQ$), if $dQ$ falls below the selected threshold, it is excluded from the charge calorimetry of the event.

The simulated energy deposited in light calorimetry, $dL/dx$, is defined as:
\begin{equation}\label{eq:dL}
\frac{{dL}}{{dx}} = \frac{{dE}}{{dx}} - \frac{{dQ}}{{dx}}
\end{equation}
where dQ comes from Eq.~\ref{eq:dQ}. The number of VUV photons is then simulated based on dL and the assumed LY. We consider several benchmark detector average light yields ($\overline{\text{LY}}$) from 35 to 220 photoelectrons per MeV of deposited energy (PE/MeV). The lower bound of 35 PE/MeV is chosen because it is close to the ($\overline{\text{LY}}$) of the DUNE vertical drift far detector under construction~\cite{DUNE_VD_TDR}. Higher $\overline{\text{LY}}$ values, up to 220 PE/MeV, are studied to explore what $\overline{\text{LY}}$ future LArTPC should aim to achieve. In LAr, approximately 21622 photons are produced per MeV of deposited energy for a minimum ionizing particle (MIP) with ${R}_{\text{c}}=$0.7 at a typical electric field of 0.5~kV/cm~\cite{Selfcompensatinglight4GeV}. The overall photon collection efficiency (PCE) is defined as the ratio between the number of detected photoelectrons in the full photon detection system and the total number of photons initially produced. For each benchmark $\overline{\text{LY}}$, the corresponding PCE is calculated as:
\begin{equation}\label{eq:PCE}
\text{PCE} (\overline{\text{LY}}) = \overline{\text{LY}}/21622
\end{equation}
The PCE values for five benchmark $\overline{\text{LY}}$ are listed in Table~\ref{tab:PCE}.

\begin{table}[tpb]
\centering
\begin{tabular}{r|c|c|c|c|c}
\hline\hline
    $\overline{\text{LY}}$ (PE/MeV)~ & ~35~ & ~100~ & ~140~ & ~180~ & ~220~  \\
    PCE~ & ~0.16\%~ & ~0.46\%~ & ~0.65\%~ & ~0.83\%~ & ~1.01\%~ \\
\hline\hline
\end{tabular}
\caption{The photon collection efficiency (PCE) of five assumed $\overline{\text{LY}}$.}
\label{tab:PCE}
\end{table}

For each individual energy deposition, the simulated expected average number of produced photons, $\overline{{N}}_{\text{ph}}$, in LAr is:
\begin{equation}\label{eq:NPE}
\overline{{N}}_{\text{ph}} = \frac{{dL}}{19.5 \ \text{eV}} \times \text{PCE} (\overline{\text{LY}})
\end{equation}
where 19.5 eV is the average energy required to produce a single 127 nm photon in LAr~\cite{IonizationfracLAr}. The detected number of photoelectrons, $N_{\text{PE}}$, is simulated by applying Poisson smearing:
\begin{equation}
{N}_{\text{PE}} = \text{Poisson}(\overline{{N}}_{\text{ph}})
\end{equation}
The final detected energy deposited in light is then derived from the detected photoelectrons for each
individual energy deposition:
\begin{equation}\label{eq:detectedlight}
{L}_{\overline{\text{LY}}} = \frac{{N}_{\text{PE}} \times 19.5 \ \text{eV}}{\text{PCE} (\overline{\text{LY}})}
\end{equation}
and summed across the event. A similar calculation taking into account the non-uniform LY and position reconstruction uncertainty is detailed in Appendix~\ref{sec:nonuniformly}.

\begin{figure}[htp]
  \centering  \includegraphics[width=0.9\columnwidth]{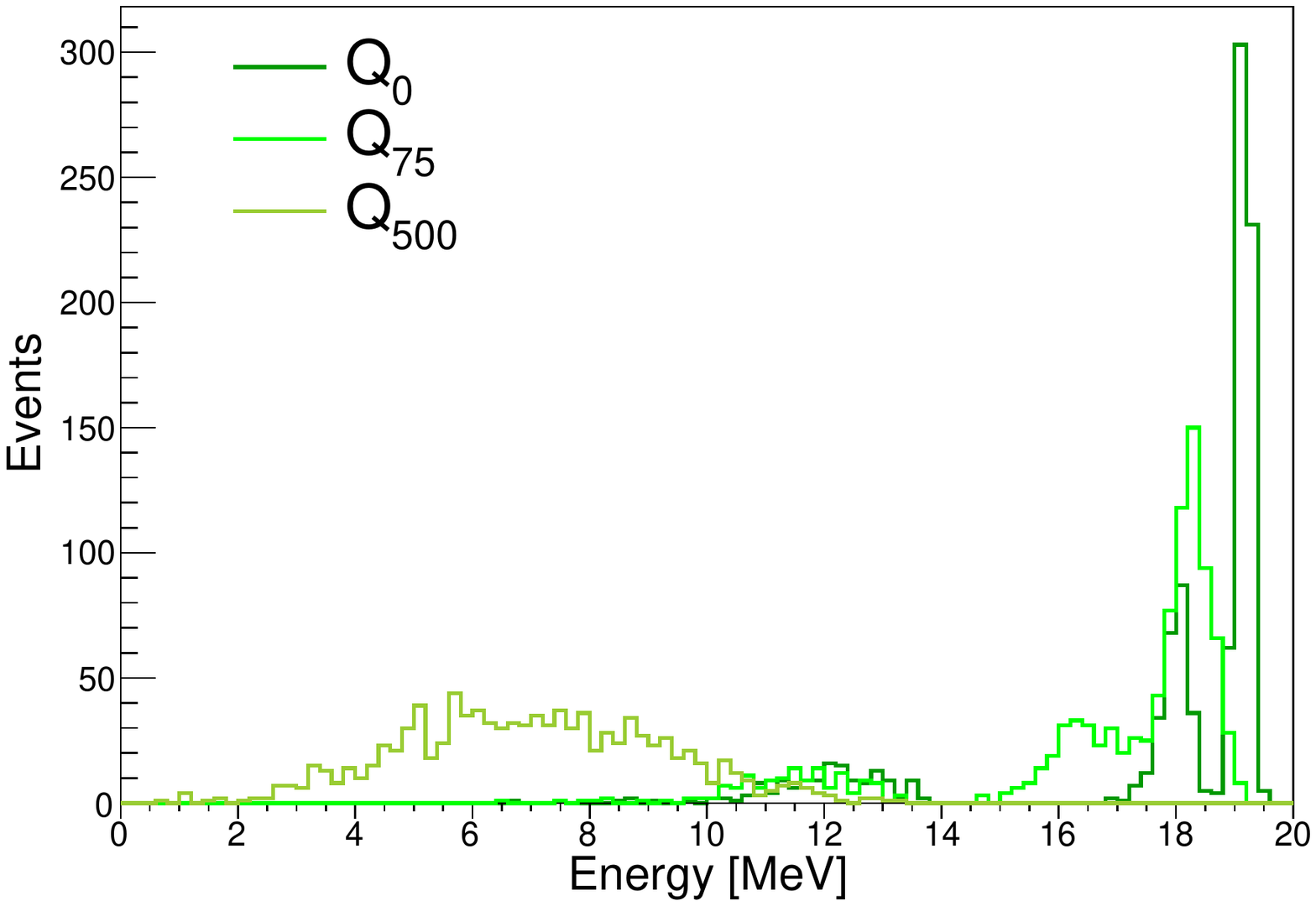}
  \centering  \includegraphics[width=0.9\columnwidth]{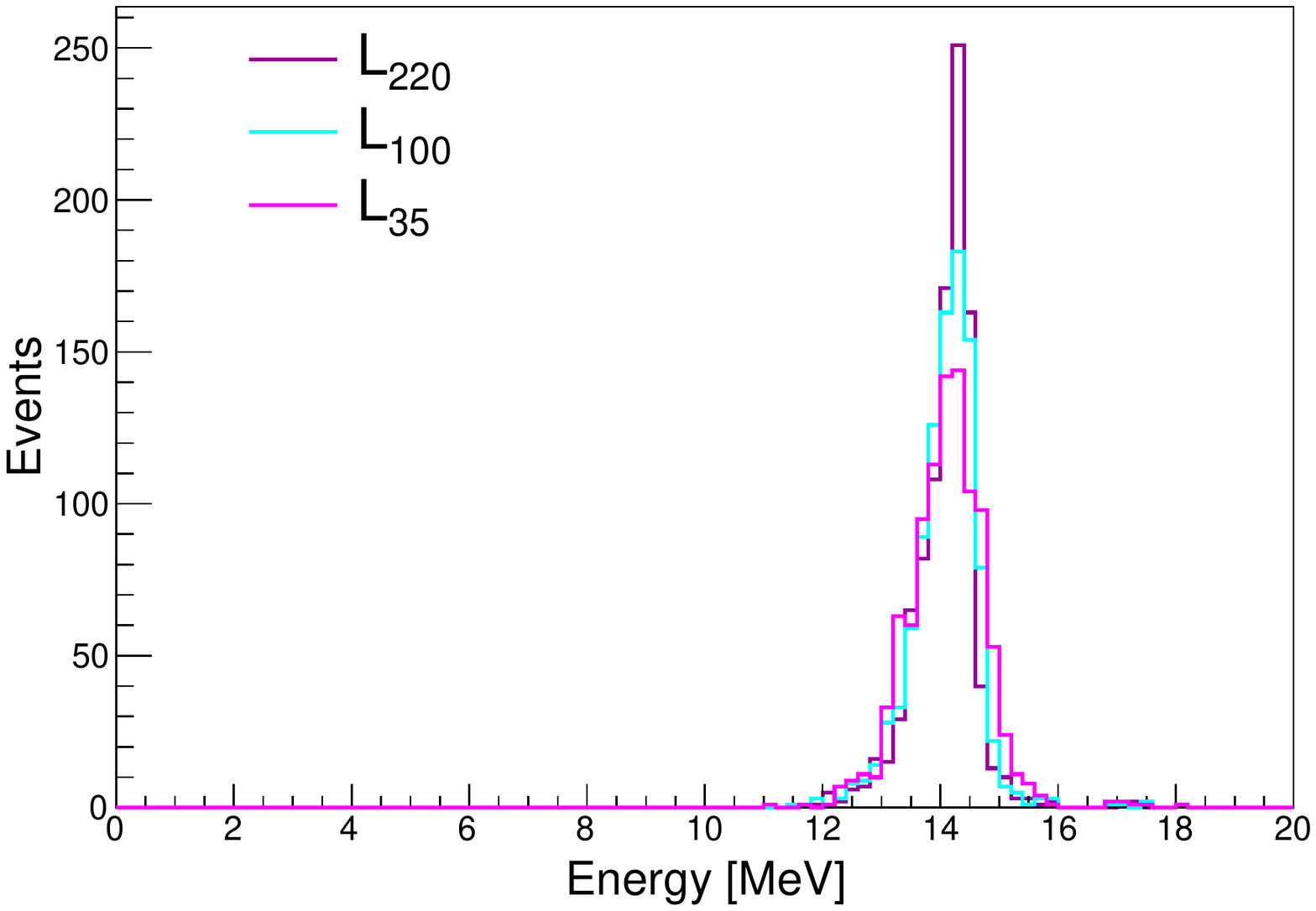}
  \centering  \includegraphics[width=0.9\columnwidth]{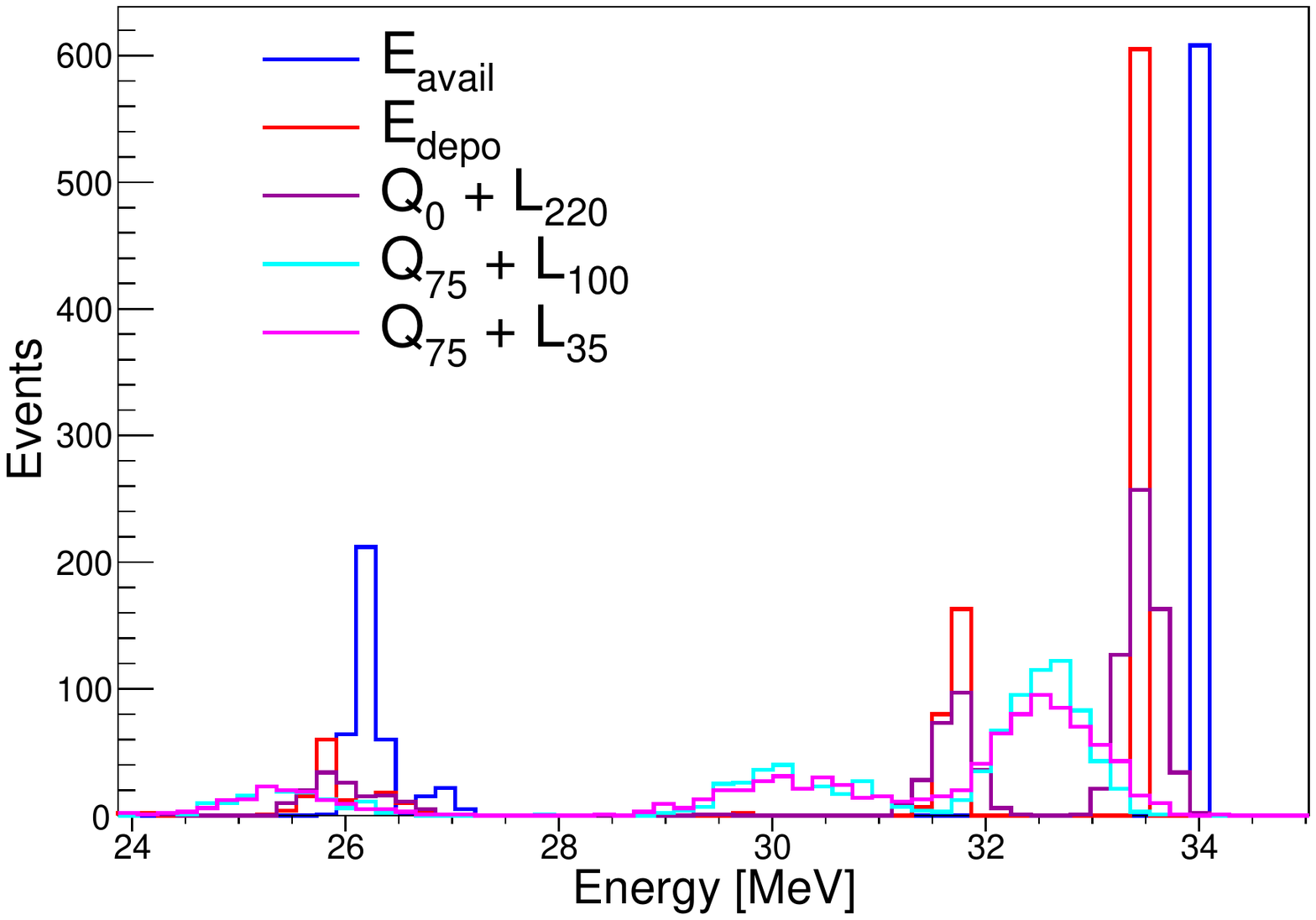}
  \caption{Deposited energy distribution for charge (top), light (middle), and combined calorimetry (bottom) for 35 MeV $\nu_{e}$-Ar CC events. In the legend, the energy deposited as charge calorimetry at zero, 75 keV, and 500 keV detection thresholds are denoted as ${Q}_{0}$, ${Q}_{75}$, and ${Q}_{500}$. The energy deposited as light calorimetry under a uniform $\overline{\text{LY}}$ of 220 PE/MeV, 100 PE/MeV, and 35 PE/MeV are denoted as ${L}_{220}$,  ${L}_{100}$, and ${L}_{35}$.}
  \label{fig:DetectedQL}
\end{figure}

Fig.~\ref{fig:DetectedQL} shows the distributions of deposited charge and light from 35 MeV $\nu_{e}$ CC interactions on Ar. In addition to ${E}_{\text{avail}}$ and ${E}_{\text{depo}}$ described in Sec.~\ref{sec:sim}, the energy deposited as charge at zero charge detection threshold, ${Q}_{0}$, primarily peaks around 19 MeV (Fig.~\ref{fig:DetectedQL} top plot). Secondary peaks in the ${Q}_{0}$ distribution originate from ${E}_{\text{depo}}$ due to events with knockout hadrons. The secondary peak at 18 MeV in ${Q}_{0}$ corresponds to events with neutron emission, while the broader peak at 12 MeV consists of events with a proton or an $\alpha$ particle. Compared to ${E}_{\text{depo}}$, the Birks model introduces additional broadening in ${Q}_{0}$. This broadening is more pronounced for secondary peaks because knockout hadrons are not MIPs, leading to greater variation in recombination effects as described by the Birks model. 

Applying a charge detection threshold of 75 keV further broadens the deposited charge calorimetry, ${Q}_{75}$, relative to ${Q}_{0}$, and introduces a negative energy bias. The 500 keV threshold significantly smears the charge calorimetry distribution, ${Q}_{500}$, reducing the precision of energy reconstruction. 

\begin{figure}[ht]
  \centering  \includegraphics[width=0.9\columnwidth]{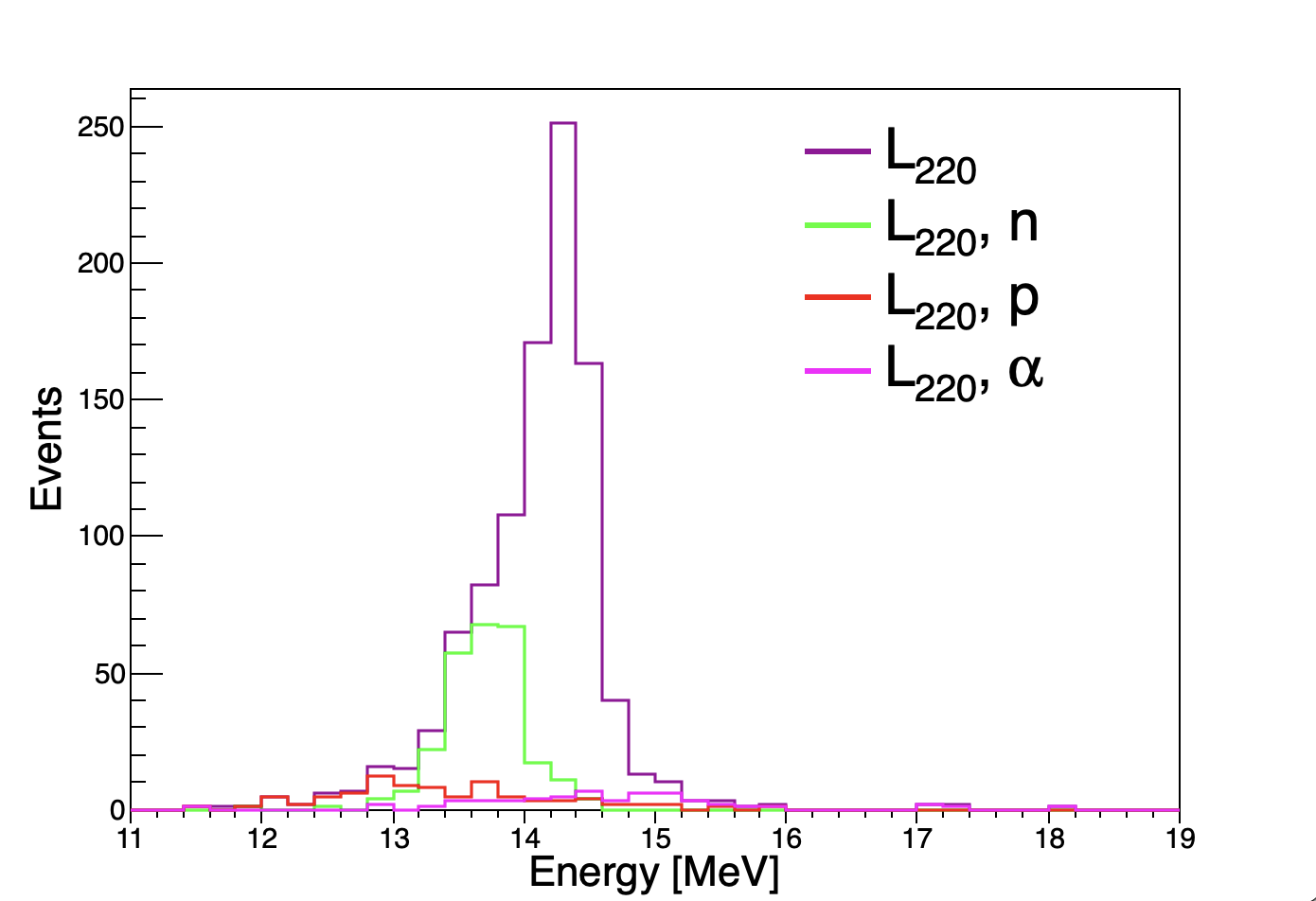}
  \caption{Deposited energy in light assuming $\overline{\text{LY}}$ = 220 PE/MeV for 35 MeV $\nu_{e}$-Ar CC events (purple), overlaid with distribution of events with knockout neutrons (green), protons (red), and $\alpha$s (pink).}
  \label{fig:dL_hadron}
\end{figure}

For energy deposited as light, the events are centered around 14 MeV (Fig.~\ref{fig:DetectedQL} middle plot). The secondary peaks are less distinct and merge with the primary peak. This effect is further illustrated in Fig.~\ref{fig:dL_hadron}, which shows the energy deposited in light for $\overline{\text{LY}}$ = 220 PE/MeV (${L}_{220}$). Specifically, for events with neutrons, the net bias in deposited energy is $-$1.8 MeV. The total deposited energy of these events typically fall within the two-standard-deviation band of the primary light peak. Consequently, neutron emission events populate the lower tail of the $\text{L}_{220}$ distribution in Fig.~\ref{fig:dL_hadron} rather than forming a distinct secondary peak, as observed in the charge distribution. 

It is important to note that nuclear quenching effects in hadron-nucleus interactions are not explicitly simulated in this study. In particular, a lower scintillation efficiency has been observed in tens of keV nuclear recoils in LXe~\cite{Xe_nuclearquenching}.  For $\alpha$ particles, a smaller light production is observed possibly explained by a biexcitonic quenching process; while for few MeV protons, this was calculated to be negligible~\cite{Nuclear_quenching}. For neutrons in this study, the potential extra light loss could shift neutron emission events further from the primary peak in the light distribution. 

The single peak structure of the light distribution is observed across all $\nu_{e}$ samples from 5 to 50 MeV and holds true for all benchmark $\overline{\text{LY}}$. Unlike the charge calorimetry, no systematic energy bias is introduced when varying $\overline{\text{LY}}$. However, energy smearing effects become more pronounced as $\overline{\text{LY}}$ decreases from 220 PE/MeV to 35 PE/MeV. In Fig.~\ref{fig:DetectedQL} middle plot, the standard deviations are 0.56 MeV, 0.59 MeV, and 0.67 MeV for ${L}_{220}$,  ${L}_{100}$, and ${L}_{35}$ respectively.

Finally, the event ${E}_{\text{depo}}$ is largely recovered when combining charge and light calorimetry, ${Q}_{0}$ + ${L}_{220}$ (Fig.~\ref{fig:DetectedQL} bottom plot). Similar to charge-only calorimetry, the application of a 75 keV charge detection threshold introduces spectral broadening and an energy bias in the combined calorimetry. Additionally, implementing the modified Box model\cite{ModifiedBoxModel} to simulate recombination effects yields similar results to those obtained using the Birks model.

\section{Capabilities from Enhanced Light Detection}
\label{sec:ereco}

\subsection{Energy Reconstruction}
\label{sec:QLreco}

In this section, we examine the potential for energy reconstruction using charge and light calorimetry in a future LArTPC with an enhanced photon detection system. The ${E}_{\text{avail}}$ for $\nu_{e}$ CC events can be reconstructed by applying a scaling factor to the deposited energy. The scaling factor represents the fraction of ${E}_{\text{avail}}$ that is converted into deposited energy in charge, light, or combined calorimetry in each event. As an example, we consider charge calorimetry with a detection threshold of 75 keV, ${Q}_{75}$. From simulation, we determine a scale factor of 0.54, taken as the mode of the ratio distribution between ${Q}_{75}$ and ${E}_{\text{avail}}$. The reconstructed energy from ${Q}_{75}$ is then given by:
\begin{equation}\label{eq:ErecoQ}
{E}_{\text{reco}, \ {Q}_{75}} = {Q}_{75}/0.54
\end{equation}
Similarly, the event energy can be reconstructed using the total deposited energy in light, for instance, with an assumed $\overline{\text{LY}}$ = 220 PE/MeV:
\begin{equation}\label{eq:ErecoL}
{E}_{\text{reco}, \ {L}_{220}} = {L}_{220}/0.42
\end{equation}
Alternatively, the $\nu_{e}$ event energy can be reconstructed by combining charge and light calorimetry:
\begin{equation}\label{eq:ErecoQL}
{E}_{\text{reco}, \ {Q}_{75}\ +\ {L}_{220}} = ({Q}_{75} + {L}_{220})/0.96
\end{equation}

Below we show three analyses (analysis A, B, and C) that go from the most basic to the most difficult scenario regarding the treatment of hadrons. The energy reconstruction capability of the light system is demonstrated through these analyses. We also note that low-energy radioactive backgrounds are not considered in any of the three analyses. Identifying signal-related energy depositions across the detector volume in a high-multiplicity, low-energy environment could present significant challenges if the rate of background activities is very high. Background energy deposits from internal or external radiological sources may be misidentified or mis-clustered as signal deposits. A detailed simulation incorporating both signal and radiological backgrounds is necessary to evaluate the detector’s ability to separate events and assess physics sensitivity, using both charge and light detection. However, such a study is beyond the scope of this paper and warrants a separate investigation. In the analysis presented below, we assume that signal-related energy depositions are efficiently identified throughout the detector volume, with effective charge-light matching.

\subsubsection{Analysis A - Exclude energy deposits from neutron capture}
\label{sec:analysisA}
In a realistic LArTPC, there will be inefficiency at tagging neutron captures since they happen much later (more than one drift time) than the initial neutrino interaction. It is straightforward for physics analysis not to include energy deposits from neutron captures. Therefore, for the most basic analysis, we exclude all energy deposits from neutron captures. 

The deposited energy distribution after excluding neutron energy deposits is shown in Fig.~\ref{fig:DetectedQL_no_ncap_depo} overlaid with some distributions from Fig.~\ref{fig:DetectedQL} for comparison. For the total deposited energy (Fig.~\ref{fig:DetectedQL_no_ncap_depo} top plot), the previous distinct neutron emission events now cluster  close with the other hadron emission events around 26 MeV. After taking into account energy partition based on Birks model, the neutron emission events now peak at around 14.5 MeV in charge calorimetry, as shown in the middle plot of Fig~\ref{fig:DetectedQL_no_ncap_depo}. Finally, due to the loss of the capture energy deposits, the neutron emission events cluster at a distinct secondary peak at around 11 MeV in the light calorimetry in contrast to Fig.~\ref{fig:dL_hadron}.

\begin{figure}[htp]
  \centering  \includegraphics[width=0.9\columnwidth]{Fig3_Edep_no_ncap_edeptotal.pdf}
  \centering  \includegraphics[width=0.9\columnwidth]{Fig3_Q_no_ncap_edeptotal.pdf}
  \centering  \includegraphics[width=0.9\columnwidth]{Fig3_L_no_ncap_edeptotal.pdf}
  \caption{Deposited energy distribution for the 35 MeV $\nu_{e}$-Ar CC sample after excluding energy deposits from neutron captures, overlaid with the relevant inclusive energy distribution in Fig.~\ref{fig:DetectedQL}. Top: total deposited energy. Middle: energy deposited into charge calorimetry ${Q}_{0}$. Bottom: energy deposited into light calorimetry ${L}_{220}$.}
  \label{fig:DetectedQL_no_ncap_depo}
\end{figure}

The energy distributions obtained using three reconstruction methods in Eq.~\ref{eq:ErecoQ}, Eq.~\ref{eq:ErecoL}, and Eq.~\ref{eq:ErecoQL} for 35 MeV $\nu_{e}$-Ar CC events are shown in Fig.~\ref{fig:RecoE_no_ncap_edep}. Notably, the primary peak at 34 MeV in the combined charge and light energy reconstruction (Eq.~\ref{eq:ErecoQL}) is narrower than charge-only reconstruction (Eq.~\ref{eq:ErecoQ}) or light-only reconstruction. This comes from the intrinsic anti-correlation between charge and light signals, which is assumed in the simulation model used in this study. This anti-correlation was previously observed in liquid xenon (LXe) experiments~\cite{EXO200, Xenon100}. The excellent energy resolution achieved by the combined calorimetry for events without knockout hadrons is further detailed in Appendix~\ref{sec:Eresnohadrons}.

\begin{figure}[htp]
  \centering  \includegraphics[width=0.9\columnwidth]{recoE_noncap.pdf}
  \caption{Reconstructed ${E}_{\text{avail}}$ for 35 MeV $\nu_{e}$-Ar CC events using charge only (green), light only (purple), and combined charge and light energy deposits (red). The neutron capture related energy deposits are excluded during reconstruction.}
  \label{fig:RecoE_no_ncap_edep}
\end{figure}

Energy reconstruction using the three methods is performed for all simulated samples from 5 MeV to 50 MeV under the optimistic 75 keV charge detection threshold and across all benchmark $\overline{\text{LY}}$. Since all distributions are non-Gaussian and exhibit distinct secondary peaks, we use the histogram standard deviation ($\sigma_{h}$) and mean ($\bar{E}_{h}$) to characterize the energy smearing and calculate the energy resolution. 

Fig.~\ref{fig:RecoEresNonfit_no_ncap} shows the energy resolution as a function of true ${E}_{\nu_{e}}$ and as a function of $\overline{\text{LY}}$ for the 20 MeV $\nu_{e}$ sample, using the full statistics of the simulated dataset. For $\nu_{e}$ energies below 15 MeV, the combined calorimetry method provides the best energy resolution among the three reconstruction approaches for any benchmark $\overline{\text{LY}}$. This improvement is attributed to the intrinsic anti-correlation between charge and light signals, as demonstrated in Appendix~\ref{sec:Eresnohadrons}. 

As true ${E}_{\nu_{e}}$ increases above 10 MeV, the fraction of hadron emission events rises (Fig.~\ref{fig:2DEavailEtrue} bottom plot), leading to a larger standard deviation in the reconstructed energy for all three methods and causing a degradation in energy resolution. However, above a certain energy, the undetectable binding energy from hadron emission becomes relatively small compared to the total neutrino energy, and the energy resolution restores the monotonic improvement as energy increases. For the charge-only reconstruction, the turning point is around true ${E}_{\nu_{e}}$ 25 MeV. While for the light-only reconstruction, it happens slightly earlier at around true ${E}_{\nu_{e}}$ 15 MeV. Surprisingly, the light-only reconstruction also provides the best energy resolution above 15 MeV; below 15 MeV, its performance is comparable with the combined calorimetry reconstruction.

For the 20 MeV $\nu_{e}$ sample shown in the bottom plot of Fig.~\ref{fig:RecoEresNonfit_no_ncap}, the energy resolution of the light-only method improves as $\overline{\text{LY}}$ increases from 35 PE/MeV to 100 PE/MeV. It also outperforms both charge-only and combined calorimetry reconstruction across all benchmark $\overline{\text{LY}}$. The combined calorimetry method does not show significant improvement in energy resolution at higher $\overline{\text{LY}}$, as its performance remains dominated by charge calorimetry. This study demonstrates that, given sufficient light collection, light-only reconstruction can achieve superior energy resolution compared to traditional charge-based calorimetry. 

\begin{figure}[htp]
  \centering  \includegraphics[width=0.9\columnwidth]{Eres_histrmsmean_no_ncap.pdf}
  \centering  \includegraphics[width=0.9\columnwidth]{20MeVEresVsLY_histrmsmean_no_ncap.pdf}
  \caption{Reconstructed energy resolution for all events in each $\nu_{e}$ energy sample after excluding energy deposits from neutron captures. Top: reconstructed event energy resolution as a function of true ${E}_{\nu_{e}}$ from 5 to 50 MeV. Bottom: reconstructed energy resolution as a function of benchmark $\overline{\text{LY}}$ for 20 MeV $\nu_{e}$.}
  \label{fig:RecoEresNonfit_no_ncap}
\end{figure}


\subsubsection{Analysis B - Include energy deposits from neutron capture}
\label{sec:analysisB}
\begin{figure}[htp]
  \centering    \includegraphics[width=0.95\columnwidth]{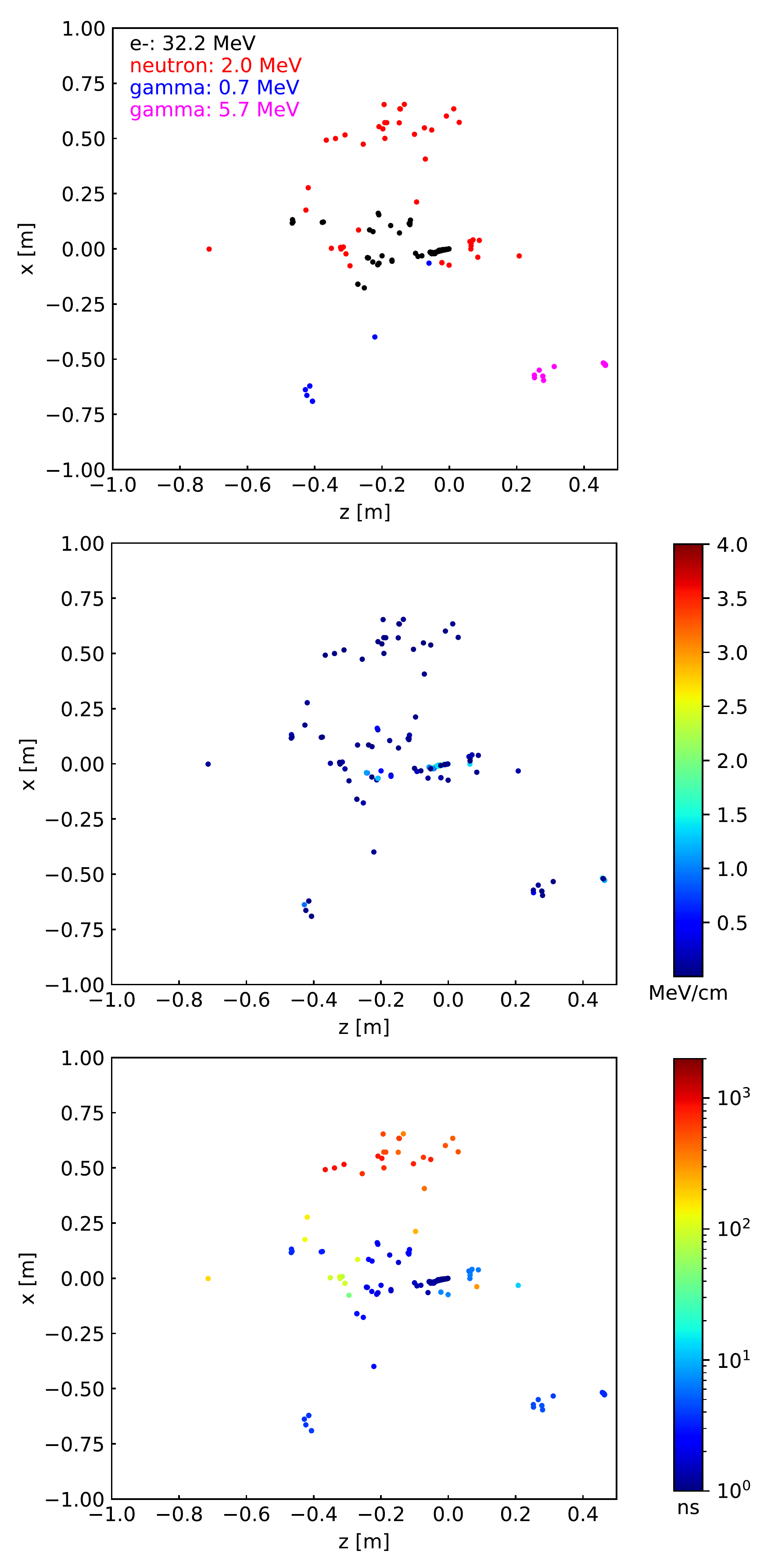}
  \caption{Displays of simulated energy deposits from a 50 MeV $\nu_{e}$-Ar CC event. The event vertex is located at the origin (0, 0, 0). The energy deposits are color-coded based on final state particles (top), linear energy transfer dE/dx (middle), and energy deposit time (bottom) respectively. In this event, the neutron has a first inelastic hadronic scattering and two gammas with energy 1.48 MeV and 0.19 MeV are produced. The neutron then goes through a series of elastic hadronic scatterings followed by a final capture on $^{40}$Ar at 1.3 $\mu$s later relative to the initial neutrino interaction. At the capture, three gammas with energy 1.92 MeV, 4.10 MeV, and 0.17 MeV are released. A 75 keV charge calorimetry detection threshold is applied in all plots.}
  \label{fig:EvtDisplay}
\end{figure}

Beyond the improvements in energy reconstruction achieved through light calorimetry and dual calorimetry, further enhancement in energy resolution is expected from the ability to tag delayed energy deposits in future LArTPCs with advanced light detection systems. In particular, tagging delayed energy deposits from neutrons and their captures using timing information from the photon detection system can significantly reduce energy smearing. An example of a simulated $\nu_{e}$-Ar CC event display is shown in Fig.~\ref{fig:EvtDisplay}. While the variation in dE/dx is relatively small across all energy deposits, the timing domain reveals key distinguishing features. Fast timing capabilities and extensive photodetector coverage in LArTPCs can help separate energy deposits from electromagnetic activities (occurring within 10 ns) and neutron scattering activities (which extend up to a few $\mu$s). In the case of neutron capture, the capture time is primarily determined by capture on the most abundant $^{40}$Ar isotope and is expected to be on the order of $\mathcal{O}$(100) $\mu$s. The energy from the cascade gamma-rays adds up to 6.1 MeV and can be identified by the photon detection system and correlated with the primary interaction.

\begin{figure}[htp]
  \centering  \includegraphics[width=0.9\columnwidth]{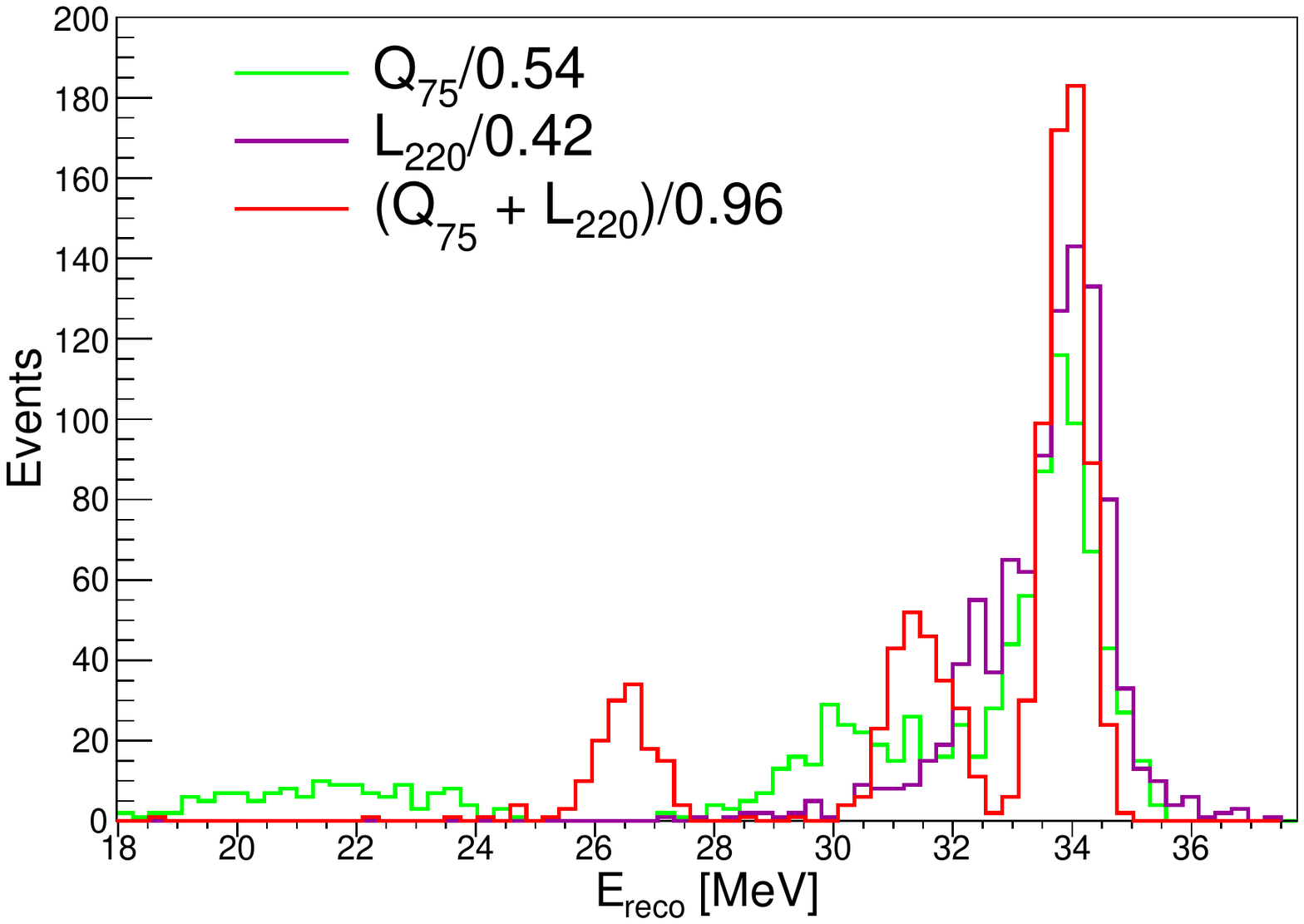}
  \caption{Reconstructed ${E}_{\text{avail}}$ for 35 MeV $\nu_{e}$-Ar CC events using charge only (green), light only (purple), and combined charge and light energy deposits (red).}
  \label{fig:RecoE}
\end{figure}

In this second analysis, we assume neutron capture associated delayed energy deposits can be included in the physics analysis. The energy distributions obtained with this assumption with the three reconstruction methods for 35 MeV $\nu_{e}$-Ar CC events are shown in Fig.~\ref{fig:RecoE}. The main difference to Fig.~\ref{fig:RecoE_no_ncap_edep} is the relocation of the neutron emission events closer to the primary peak, as hinted already by Fig.~\ref{fig:DetectedQL_no_ncap_depo}.

\begin{figure}[htp]
  \centering  \includegraphics[width=0.9\columnwidth]{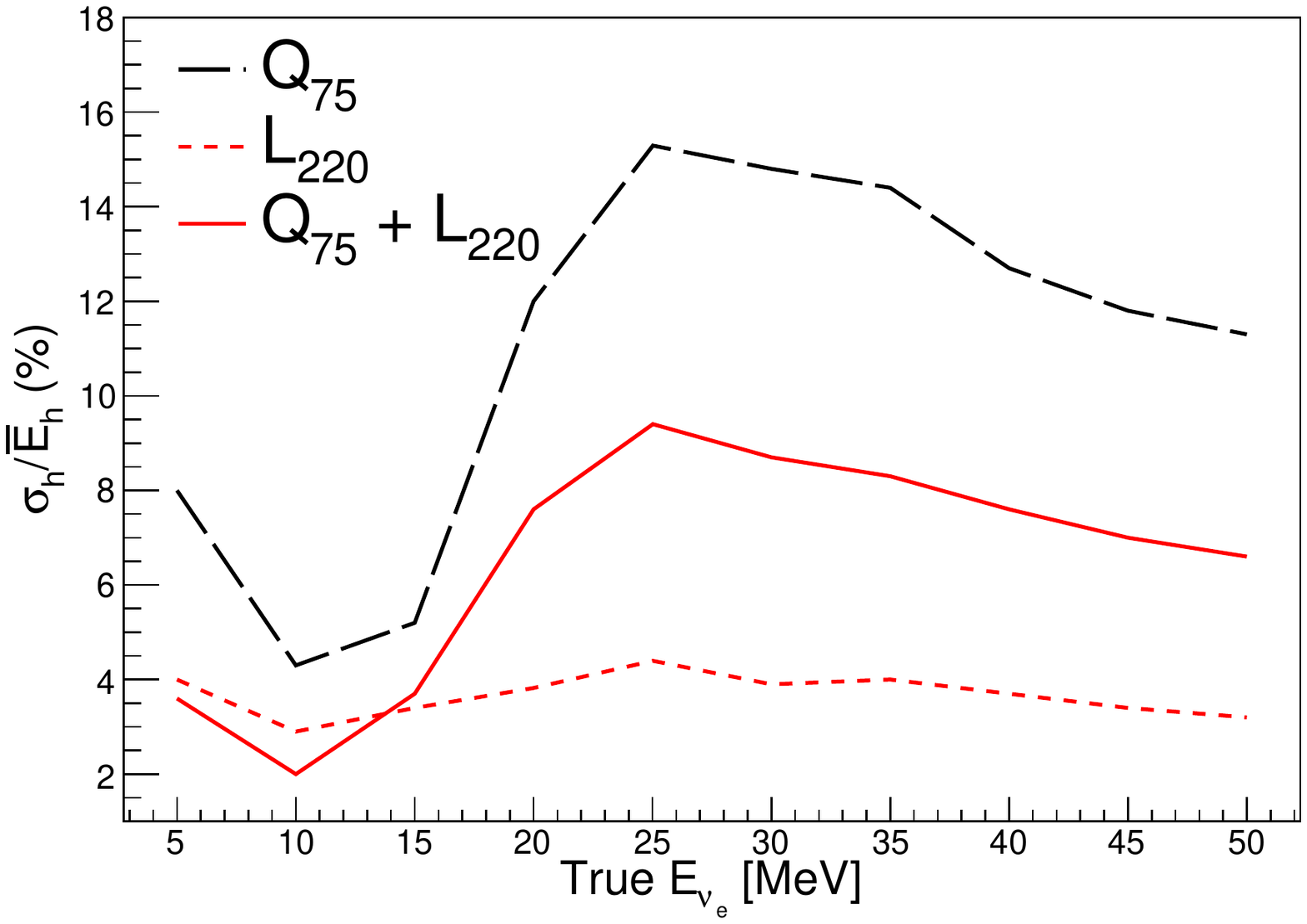}
  \centering  \includegraphics[width=0.9\columnwidth]{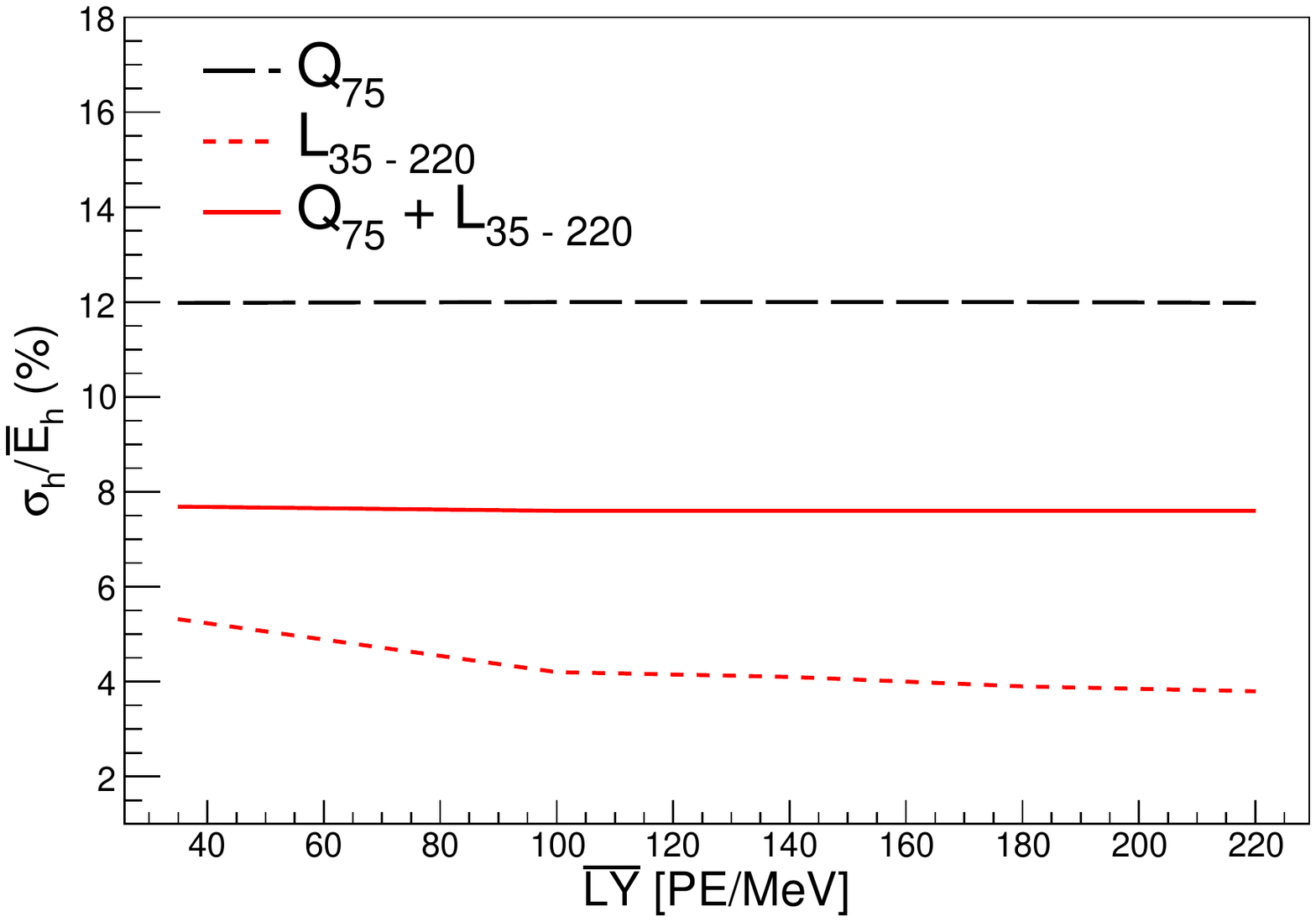}
  \caption{Reconstructed energy resolution for all events in each $\nu_{e}$ energy sample. Top: reconstructed event energy resolution as a function of true ${E}_{\nu_{e}}$ from 5 to 50 MeV. Bottom: reconstructed energy resolution as a function of benchmark $\overline{\text{LY}}$ for 20 MeV $\nu_{e}$.}
  \label{fig:RecoEresNonfit}
\end{figure}

Fig.~\ref{fig:RecoEresNonfit} shows the updated energy resolution under the assumption neutron capture released energy deposits can be included. For $\nu_{e}$ energies below 15 MeV, the combined calorimetry method still provides the best energy resolution among the three reconstruction approaches for any benchmark $\overline{\text{LY}}$. The similar degradation in energy resolution is observed starting at 10 MeV due to the onset of hadron production, followed by a recovery at around 25 MeV. One major observed difference to the case of excluding neutron capture energy deposits in Fig.~\ref{fig:RecoEresNonfit_no_ncap} is the light-only reconstruction is much less affected by hadron emission events, maintaining a nearly flat energy resolution at $\sim$4\% as a function of true ${E}_{\nu_{e}}$. This stability arises from the absence of secondary peaks, as shown in Fig.~\ref{fig:dL_hadron}. Overall the absolute energy resolution are also better than that in Fig.~\ref{fig:RecoEresNonfit_no_ncap} for each calorimetric reconstruction; the light-only reconstruction improved the most from $\sim$13\% to $\sim$4\% at true ${E}_{\nu_{e}}$ of 25 MeV. 

For $\nu_{e}$ energies above 15 MeV, light-only reconstruction provides the best energy resolution. Although not explicitly shown here,  this trend is consistent across all five benchmark $\overline{\text{LY}}$ scenarios for $\nu_{e}$ energies above 20 MeV. While combined calorimetry still improves the energy resolution relative to charge-only reconstruction, it largely inherits the resolution behavior of the charge-only method.

For the 20 MeV $\nu_{e}$ sample shown in the bottom plot of Fig.~\ref{fig:RecoEresNonfit}, the energy resolution of the light-only method improves as $\overline{\text{LY}}$ increases. It also outperforms both charge-only and combined calorimetry reconstruction across all benchmark $\overline{\text{LY}}$, a trend that holds for $\nu_{e}$ energies above 15 MeV. The combined calorimetry method does not show significant improvement in energy resolution at higher $\overline{\text{LY}}$, as its performance remains dominated by charge calorimetry.  

\subsubsection{Analysis C - Tag all hadrons and correct for the lost binding energy}
\label{sec:analysisC}

The previous analysis includes all (prompt + delayed) energy deposits seen in the detector and doesn't consider an explicit particle identification (PID) of hadrons. However, specific tagging of the hadrons using the similar photon detection system features described before can help correct for the lost binding energy associated with the hadron production. The identification of protons and $\alpha$s is possible by looking for highly localized collected charge since these hadrons have high dE/dx and are expected to only span $\sim$ 1 cm in the charge readout views in LArTPCs. An enhanced photon detection system helps to locate the region of interest in a three-dimensional sea of low-energy backgrounds so that a more efficient charge-light matching can be performed to identify these hadrons. It is promising that neutron captures can be tagged by a combined analysis of signals from both the charge and light detection systems in LArTPC through blips analysis and timing information~\cite{MeVNueArCC, ArgoNeuT-MeV}. However, all these need to be further developed and demonstrated, considering potential difficulties discussed in prior literatures~\cite{PhysRevD.99.036009, DUNE_FD_TDR_Physics}. This represents the most difficult analysis as it requires the PID of hadrons.

Building on analysis B in Sec.~\ref{sec:analysisB}, we assume that two common hadrons knocked out of the Ar nucleus in MeV-scale $\nu_{e}$ CC interactions---protons and $\alpha$ particles---can be tagged when their deposited charge energy exceeds the optimistic 75 keV threshold. Under this assumption, the hadron tagging efficiency is 84\% for protons and 18\% for $\alpha$ particles across all simulated neutrino events in the 5–50 MeV range. In the simulation, all neutrons are captured on $^{40}$Ar, and we assume all captured neutrons can be tagged in this study.

The decision to include or exclude events with tagged hadrons depends on the specific signal and background considerations of a given analysis. Here, we explore both possibilities. When including these tagged events in the reconstructed sample, we apply an energy correction by adding back an average binding energy of 7.35 MeV for each tagged proton or $\alpha$, and 7.9 MeV for each tagged neutron, while subtracting 6.1 MeV to account for the neutron capture on $^{40}$Ar. To exclude the hadron tagged events, they are simply discarded. 

\begin{figure}[htp]
  \centering  \includegraphics[width=0.9\columnwidth]{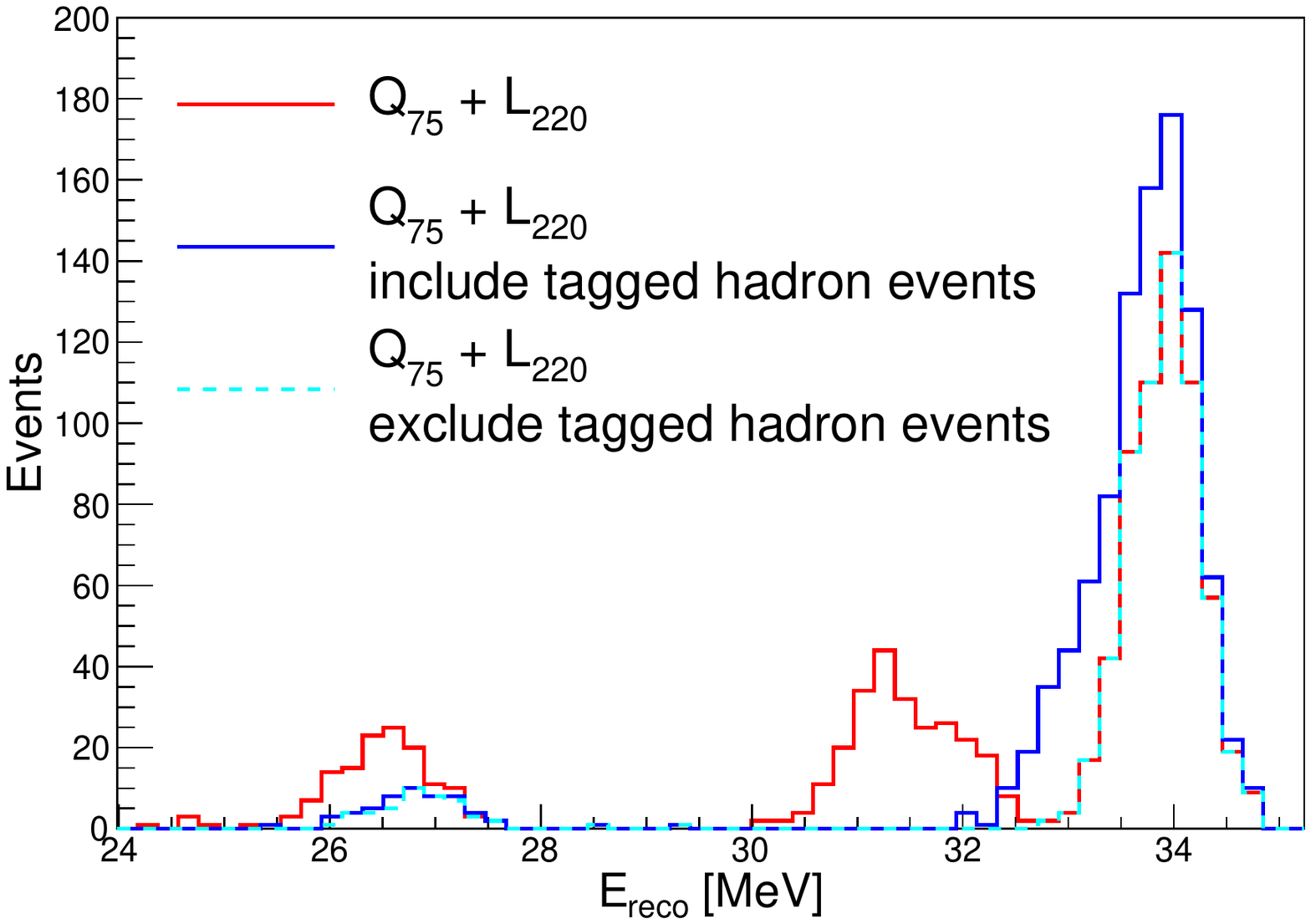}
  \caption{Reconstructed total available energy for 35 MeV $\nu_{e}$-Ar CC events with combined charge and light calorimetry without hadron tagging (red) same as Fig.~\ref{fig:RecoE}, with hadron tagging: i) tagged hadron events are energy corrected and included into the primary peak (blue); ii) tagged hadron events are excluded from the sample (cyan dashed).}
  \label{fig:nucleontagging}
\end{figure}

Fig.~\ref{fig:nucleontagging} illustrates both approaches for the 35 MeV $\nu_{e}$-Ar CC simulated sample using the combined calorimetric reconstruction from Eq.~\ref{eq:ErecoQL}. The figure clearly shows that all neutron-containing events in the 30–32 MeV range are tagged, and they can either be incorporated into the primary peak at 34 MeV or removed. Events with a proton or an $\alpha$ cluster below 28 MeV and approximately 60\% of them are tagged in the 35 MeV $\nu_{e}$ sample. The total tagging efficiency for protons and $\alpha$ particles exceeds 50\% for $\nu_{e}$ above 20 MeV.  

\begin{figure}[htp]
  \centering  \includegraphics[width=0.9\columnwidth]{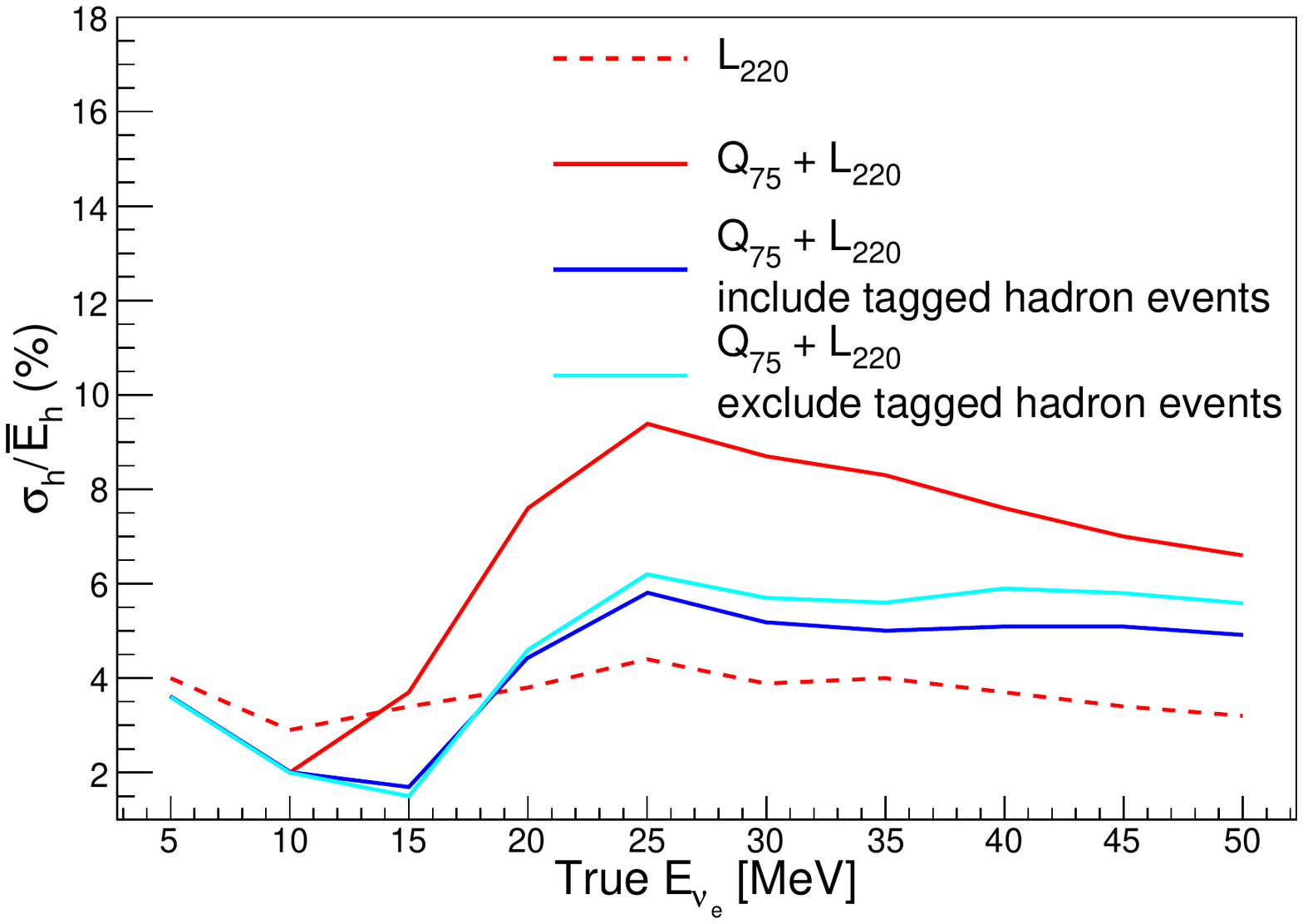}
  \centering  \includegraphics[width=0.9\columnwidth]{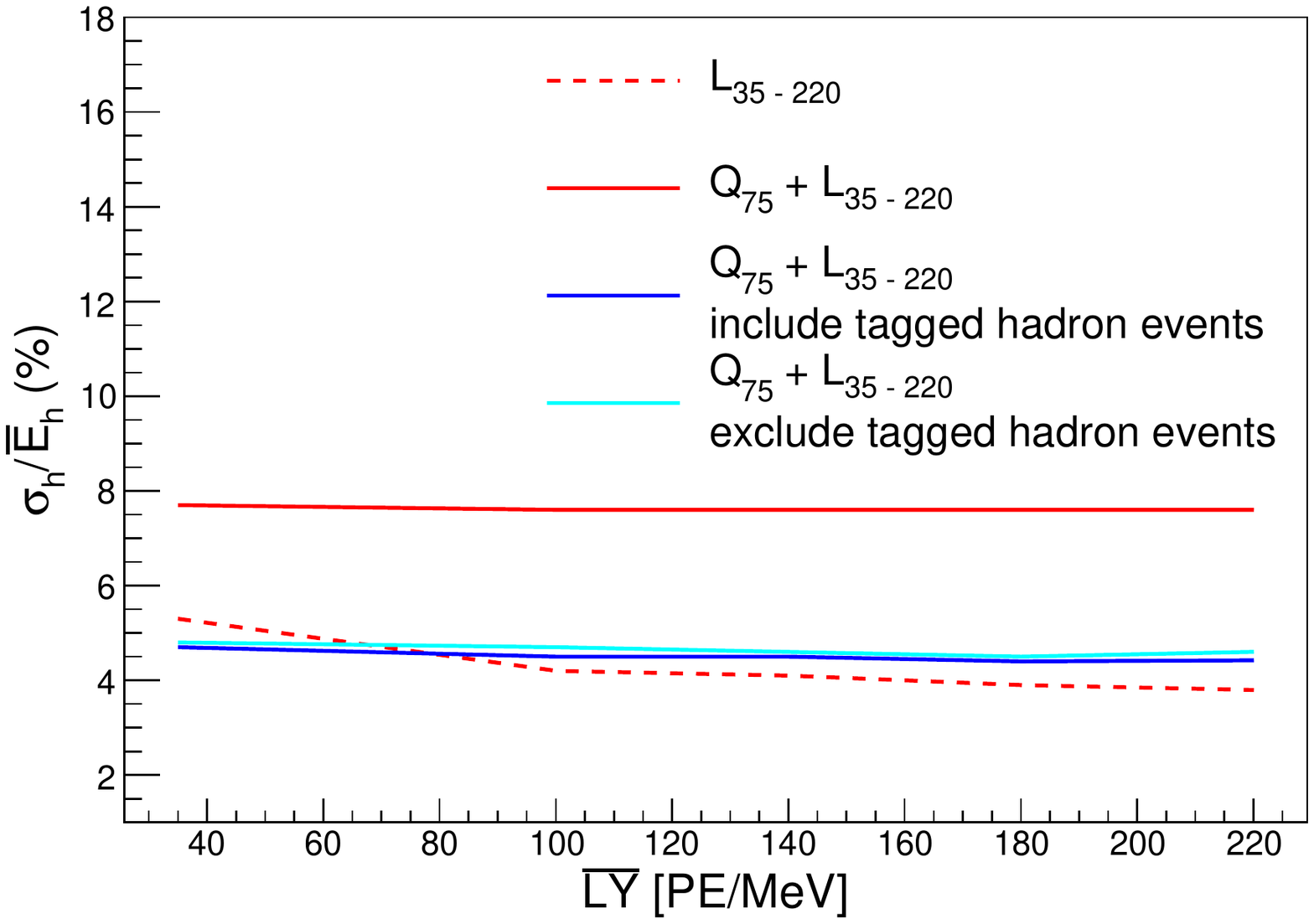}
  \caption{Reconstructed energy resolution for all events in each $\nu_{e}$ energy sample after hadron tagging. Both options of including (blue) and excluding (cyan dashed) tagged hadron events are shown. Top: reconstructed event energy resolution as a function of true ${E}_{\nu_{e}}$ from 5 to 50 MeV. Bottom: reconstructed energy resolution as a function of benchmark $\overline{\text{LY}}$ for 20 MeV $\nu_{e}$. In both plots, the light only calorimetry (red dotted line) and combined calorimetry (red solid line) without any hadron tagging shown in Fig~\ref{fig:RecoEresNonfit} are overlaid for comparison.}
  \label{fig:RecoEresNonfitPosthadrontagging}
\end{figure}

The energy resolution after hadron tagging is shown in Fig.~\ref{fig:RecoEresNonfitPosthadrontagging} for both cases: including and excluding the tagged events. Hadron tagging significantly improves the energy resolution for the combined calorimetry reconstruction. For example, in the 25 MeV sample, the energy resolution improved from 9.4\% to 5.8\%. Above 25 MeV, including tagged events results in only a marginal improvement over excluding them. Meanwhile, light-only reconstruction continues to provide the best energy resolution for all $\nu_{e}$ samples above 20 MeV. 

For the 20 MeV $\nu_{e}$ sample shown in the bottom plot of Fig.~\ref{fig:RecoEresNonfitPosthadrontagging}, post-hadron-tagging combined calorimetry achieves an energy resolution comparable to that of light-only reconstruction across all benchmark $\overline{\text{LY}}$.

We note that, according to Fig.~\ref{fig:DetectedQL} and Fig.~\ref{fig:dL_hadron}, one should avoid making the same correction to the reconstructed energy based on tagged hadrons in the light-only or charge-only energy reconstruction. This is due to large event-by-event fluctuations in $dE/dx$ for emitted hadrons, which affect the partitioning of energy between charge and light calorimetry according to the Birks model. Additionally, explicit charged PID of electrons or $\gamma$s does not further enhance the reconstructed energy resolution, as energy smearing at the deposition stage is already minimal for these particles.

The non-uniform LY is expected to have negligible impact once precise event location is reconstructed from light signals and charge-light matching, as discussed in Appendix~\ref{sec:nonuniformly}.

\subsection{Tagging Delayed Gamma in $\nu_{e}$-Ar CC Interactions}
\label{sec:gammatag}

The $\nu_{e}$-Ar CC cross section receives contributions from both Fermi and Gamow-Teller transitions~\cite{DUNE_SNB_Pointing, Marley_nueArCC, DSNB_nuclear_transitions, SNB_parametizedfit}. Approximately 65$\%$ of Fermi transitions in Ar involve delayed $\gamma$ emission due to a long-lived intermediate nuclear state of the daughter $^{40}$K$^\ast$ nucleus~\cite{FTransition}. In these interactions, the nucleus is initially excited to 4.38 MeV and subsequently deexcites via a prompt 2.73 MeV $\gamma$ emission, followed by a delayed 1.65 MeV $\gamma$ emission. The delayed 1.65 MeV $\gamma$ is associated with a characteristic decay time of 480~ns, resulting in a distinctive two-pulse timing structure in the light signal, which can be identified using the photon detection system. The reconstruction of these low energy $\gamma$s could be further strengthened by detailed blip reconstruction techniques, as demonstrated in Ref~\cite{MeVNueArCC}. 

High-$\overline{\text{LY}}$ photon detection in LArTPCs, such as that achievable with the APEX design~\cite{DUNE_Phase2}, can significantly reduce radiological background by requiring coincident activities. This is made possible by the favorable energy, vertex, and time resolution of the light signal. An application of these capabilities in solar neutrinos is possible. These studies are ongoing and will be reported later. We briefly mention potential impact here:

\textit{Improved vertex reconstruction:} with large background rates from external and internal radioactivity, matching low-energy charge and light deposits is difficult in a large LArTPC.  Vertex reconstruction is $\sim$~1~cm for charge and limited by the light signal. Increased light collection efficiency improves vertex localization which in turn will improve the matching efficiency to charge signals and facilitate the reconstruction of more MeV scale activities to the cm scale.  This is crucial for rejecting $\gamma$ backgrounds produced in the rock surrounding the LArTPC. The demonstration of the improved charge-light matching efficiency and the improved performance to reconstruct MeV scale energy deposition with such an enhanced light detection system will be reported in a future work.

\textit{Fermi transition tagging:} Separating Fermi transition interactions that emit a 1.65~MeV delayed gamma from Gamow-Teller signal and radiological background would impact solar neutrino analysis and supernova pointing in LArTPC detectors.  Searching for two flashes, possible through improved energy and vertex reconstruction, would dramatically reduce backgrounds from natural radioactivity and neutron capture for solar neutrino measurements.  For a supernova neutrino burst, these selected events would also have improved energy resolution as no nucleons are emitted in Fermi transitions.  Fermi and Gamow-Teller interactions also have different angular distributions for final state leptons: $1+\cos\theta_e$ and $1-\frac{1}{3}\cos\theta_e$ respectively~\cite{Marley_main}.  If Fermi transitions could be tagged on an event-by-event basis, their angular dependence would give additional information on pinpointing the supernova for optical followup~\cite{DUNE_SNB_Pointing}.

\section{Applications to low energy physics}
\label{sec:app}

The energy reconstruction incorporating the light calorimetry and the assumed hadron tagging capability described in Sec~\ref{sec:ereco} are applied to the DSNB search as an example physics application. To refine energy smearing, the energy granularity of the Monte Carlo samples described in Sec~\ref{sec:sim} is enhanced by adding additional mono-energetic $\nu_{e}$ events from 5 MeV to 80 MeV in 0.5 MeV increments. For each energy step, 1000 events are generated and simulated. The final energy smearing matrices are shown in Fig.~\ref{fig:response_matrix} in Appendix~\ref{sec:Esmearmatrices}.

The DSNB represents the cumulative neutrino flux from all core-collapse supernovae out to several Gpc that arrives at Earth~\cite{DSNB_flux_model}. It is a guaranteed yet undiscovered signal. LArTPC experiments offer an unique capability to constrain the $\nu_{e}$ flux, complementing other next-generation neutrino experiments such as JUNO~\cite{JUNO_DSNB} and Hyper-Kamiokande~\cite{HyperK_DSNB}, which are more sensitive to the $\overline{\nu}_{e}$ flux. The observation and measurement of DSNB would provide critical insights into the local supernova density and the relative DSNB contribution from black hole- and neutron star-forming supernovae. 

\begin{figure}[htp]
  \centering  \includegraphics[width=0.9\columnwidth]{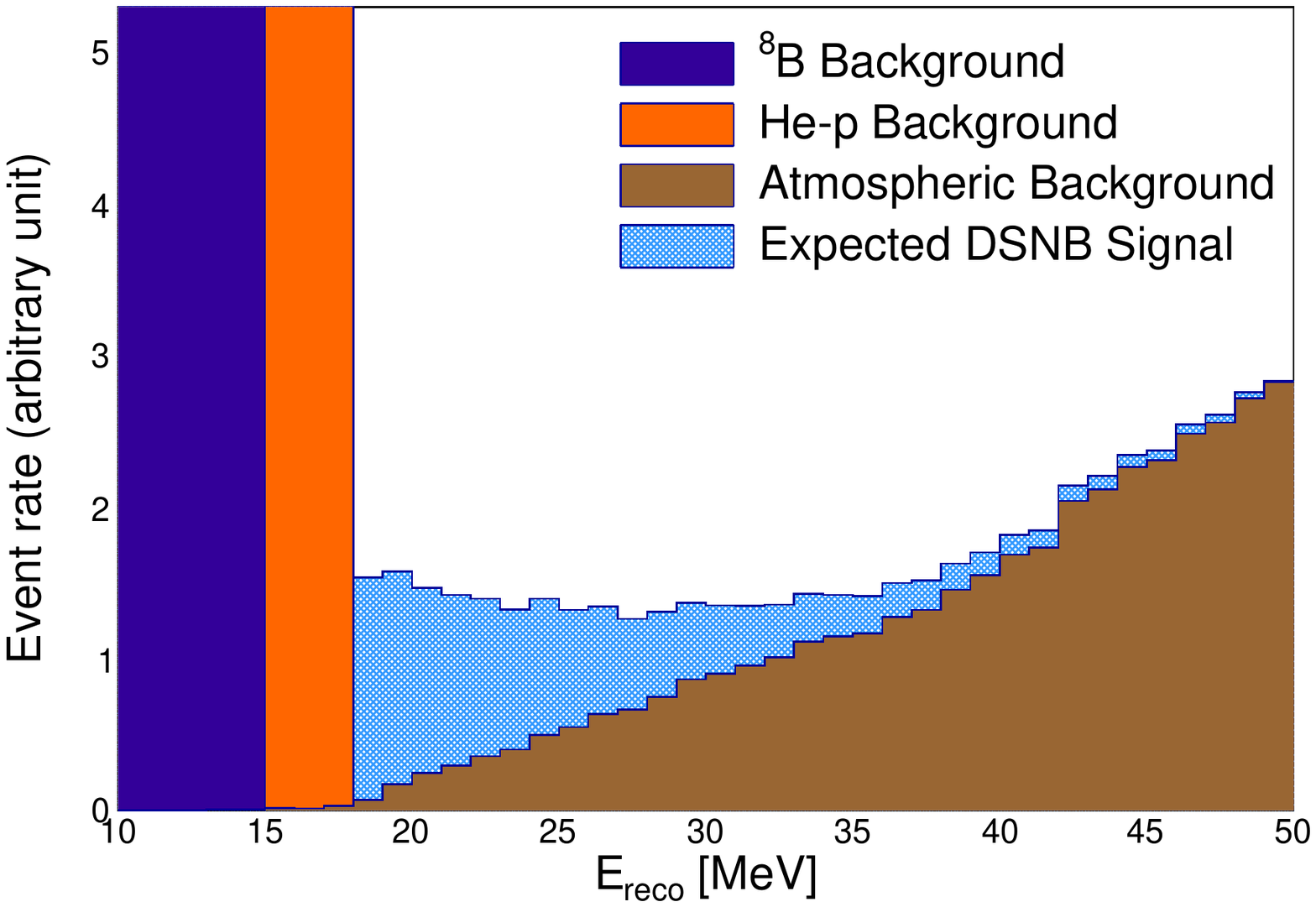}
  \caption{Expected DSNB spectra with backgrounds at 400 kt-yrs exposure under the best energy reconstruction strategy of combined calorimetry at a $\overline{\text{LY}}$ of 180 PE/MeV, ${E}_{\text{reco}, \ {Q}_{75}\ +\ {L}_{180}}$, and exclusion of hadron tagged events.}
  \label{fig:DSNBspectra}
\end{figure}

Fig.~\ref{fig:DSNBspectra} shows the expected DSNB spectra at 400 kt-yrs exposure using the ${E}_{\text{reco}, \ {Q}_{75}\ +\ {L}_{180}}$ smearing, assuming a LArTPC module with enhanced photon detection, including the hadron tagging capability as described in Sec.~\ref{sec:analysisC}. The assumptions for the DSNB $\nu_{e}$ flux model follow those in Ref.~\cite{DSNB_flux_model}. The atmospheric $\nu_{e}$ flux is taken from Ref.~\cite{SURF_atm_flux} assuming the detector is at the SURF site. The solar $\nu_{e}$ flux is derived from well-known nuclear cross-section measurements~\cite{solarflux}, with oscillations applied using parameters $\Delta {m}_{12}^2 = 4.86 \times 10^{-5} \text{eV}^2$ and $\sin^2 \theta_{12} = 0.306$. Although the $\Delta {m}^{2}_{21}$ value used is slightly outdated, its impact on the results is negligible. The $^{3}$He-p endpoint 18.8 MeV does not show up in Fig.~\ref{fig:DSNBspectra} and instead stops at 18 MeV because only the total reconstructed available energy is plotted and there is a 1 MeV binding energy change from the initial Ar nucleus to the K nucleus in the MARLEY event generator, as explained in Sec~\ref{sec:sim}. The solar neutrino backgrounds in Fig.~\ref{fig:DSNBspectra} are actual shapes from the flux model after energy smearing and are part of the stacked histogram. They appear band-like because the energy resolution is very good ($\sim$1.3\% at 18 MeV).

The DSNB signal is primarily constrained to a narrow energy region of interest (ROI). At the lower end of this ROI, the dominant background comes from the intense solar $^{3}$He-p and $^{8}$B $\nu_{e}$ fluxes, while at the higher end, atmospheric neutrino backgrounds gradually increase. Notably, the atmospheric $\nu_{e}$ flux below $\sim$ 100 MeV remains uncertain, making sensitivity optimization in the lower-energy bins within the ROI crucial.

For solar backgrounds, energy smearing caused by the secondary peak from hadron knockout effects (as described in Sec~\ref{sec:analysisC}) does not impact the DSNB search. The primary reconstructed energy peak from combined charge and light calorimetry, shown in Fig.~\ref{fig:RecoEresFit} in Sec~\ref{sec:QLreco}, provides the best opportunity to improve the search sensitivity. Conversely, for atmospheric background at the high-energy bins of the DSNB ROI, the secondary peak does affect the search sensitivity. Therefore, tagging $\nu_{e}$~CC events with knockout hadrons is crucial to mitigating smearing effects. Alternatively, one can rely on light-only energy reconstruction, which achieves better energy resolution above 20 MeV than combined calorimetry post hadron tagging, as demonstrated in Fig.~\ref{fig:RecoEresNonfitPosthadrontagging}.

To evaluate the impact of different reconstruction approaches, we apply energy smearing matrices from five reconstruction strategies: charge-only, light-only, combined calorimetry, and combined calorimetry with and without tagged hadron events, using all benchmark $\overline{\text{LY}}$. As discussed in Appendix~\ref{sec:nonuniformly}, the effect of non-uniform LY can be corrected with a precise knowledge of the event position, the corresponding energy smearing is expected to remain the same as the relevant light reconstruction considered above. So we don’t apply energy smearing under non-uniform light collection here.

To compare the performance of different energy reconstruction methods, the -2\text{log-likelihood} of the DSNB search (shown in Fig.~\ref{fig:DSNBspectra}) is defined and calculated for 400 kt-yr exposure:
\begin{equation}\label{eq:2LogLikelihood}
-2\text{log-likelihood} = 2\sum_{i} ({E}_{i}-{O}_{i} + {O}_{i}\times \log \frac{{O}_{i}}{{E}_{i}})
\end{equation}
where $i$ runs over all bins in the energy ROI, ${E}_{i}$ represents the expected background in the $i^{th}$ energy bin, and ${O}_{i}$ is the sum of signal and background in the same bin. 

\begin{figure}[htp]
  \centering  \includegraphics[width=0.9\columnwidth]{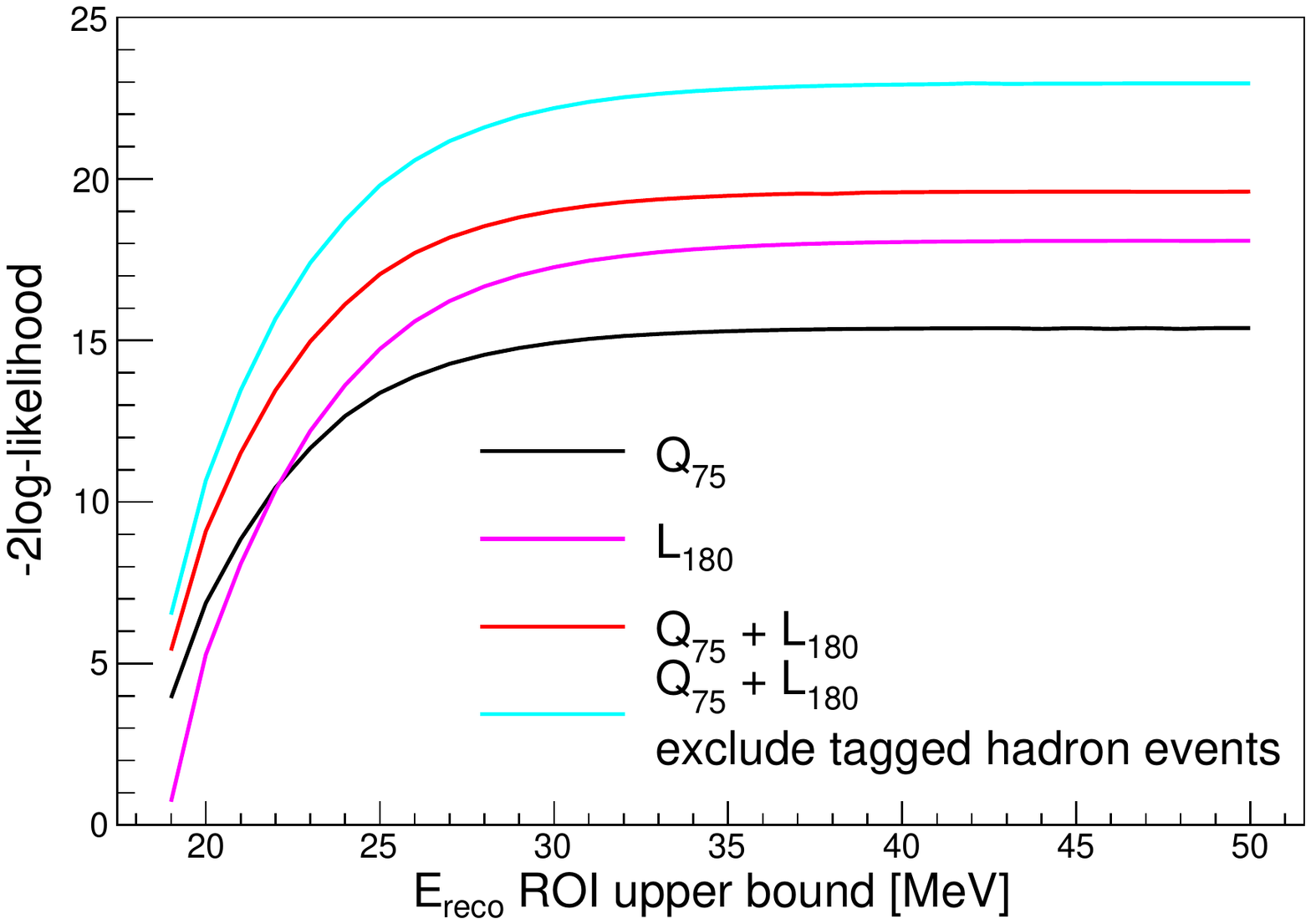}
  \caption{Expected DSNB sensitivities -2\text{log-likelihood} for 400 kt-yr exposure under different energy reconstruction strategies.}
  \label{fig:DSNBsensitivity}
\end{figure}

The best DSNB search sensitivity from different energy reconstruction strategies is shown in Fig.~\ref{fig:DSNBsensitivity} as a function of the upper bound of the energy ROI, with the lower bound fixed at 18 MeV in all cases. The light-only reconstruction, assuming a benchmark $\overline{\text{LY}}$ of 180 PE/MeV (${L}_{180}$), achieves a higher plateau sensitivity than the charge-only reconstruction, which assumes an optimistic 75 keV charge detection threshold (${Q}_{75}$). However, ${Q}_{75}$ yields better sensitivity at the lower-energy end of the ROI compared to ${L}_{180}$, as the secondary peaks in charge-only reconstruction smear the solar background to even lower energies, shifting it outside the DSNB ROI.

Energy reconstruction using combined calorimetry provides a slight improvement in plateau sensitivity. However, when combined with the hadron tagging capability described in Sec.\ref{sec:analysisC}, it achieves the highest plateau sensitivity. Hadron tagging is particularly beneficial at higher-energy ROIs, where atmospheric background dominates. The inclusion or exclusion of tagged hadron events has minimal impact on sensitivity, so only one result is presented in Fig.\ref{fig:DSNBsensitivity}. Additionally, for $\overline{\text{LY}}$ above 180 PE/MeV, no significant improvement in sensitivity is observed for light-only, combined calorimetry, or combined calorimetry with tagged hadron events, as energy resolution is no longer the limiting factor. 

Further improvements in DSNB sensitivity could be achieved by tagging solar $\nu_{e}$ backgrounds using the directionality of final-state electrons in Fermi transitions, as described in Sec~\ref{sec:gammatag}.  This would allow statistical separation between directional solar events and the isotropic DSNB signal, enabling the ROI to be extended to lower energies and thereby increasing the DSNB event rate. Additionally, the broad energy range and baseline coverage of atmospheric neutrino oscillations in future LArTPC experiments could help constrain sub-GeV atmospheric neutrino fluxes~\cite{subGeV_atmnuosc}.

\section{Summary and Outlook}
\label{sec:summary}

We studied MeV $\nu_{e}$-Ar CC interactions and simulated the energy deposit into charge and light signals. At the generator level, an energy smearing effect is observed from hadron emission whose associated binding energy is undetectable. The dominant energy smearing at the energy deposition stage in LAr comes from neutron captures on Ar nuclei. Therefore for low energy physics it is important to tag hadron multiplicities. Further hadron PID, especially the tagging of neutrons as they are the primary knockout hadrons from these interactions according to MARLEY simulation, can help precisely correct for the energy smearing effect from neutrons. Developing techniques for tagging MeV neutrons and delayed neutron captures on Ar nucleus at smaller scale LArTPCs utilizing information from both the charge and the light detection systems will be interesting.

An energy reconstruction based on light only calorimetry, or a combined calorimetry assuming an optimistic 75 keV charge detection threshold and $\overline{\text{LY}}$ of 180 PE/MeV, both show excellent energy resolution when compared to charge-only energy reconstruction. In particular, for tens of MeV $\nu_{e}$-Ar CC events, we found the energy reconstructed based on light-only calorimetry has a better resolution than combined charge and light calorimetry when hadron emission events are included. The improved energy resolution from the combined charge and light calorimetry boosts the DSNB discovery potential compared to an optimistic charge based reconstruction at the same exposure without taking into account systematic uncertainties. Once the hadron emission events are tagged, the combined calorimetry offers the best achievable DSNB sensitivity. 

In this study low energy radioactive background is not considered, but the related background energy deposit could be mis-clustered as a signal energy deposit. Detailed simulation with both signal and radiological background should be performed to understand event separation capability and physics sensitivity using both the charge and light detection.

This study offers insights into implementing energy reconstruction using combined charge and light calorimetry. According to Eq.~\ref{eq:detectedlight}, the deposited energy in light is derived from the detected photoelectrons and also the overall PCE of the LArTPC. For MeV neutrino events which have energy deposits spread over a sphere of $\sim$50 cm, the collected photoelectrons can be easily calculated by summing up detected photoelectrons from all photodetectors. In reality, the PCE of the LArTPC varies as a function of position. It's expected light calorimetry calibration at a voxel size of 50 cm can be achieved where the LY is reasonably uniform. With the precise position information reconstructed from both the charge and the light detection systems, Eq.~\ref{eq:PCE} can be used to obtain the PCE for the voxel where the MeV interaction happens. The energy deposited in charge shown in Eq.~\ref{eq:dQ} can be obtained based on calculations in Ref~\cite{Selfcompensatinglight4GeV}. Furthermore, for a homogeneous calorimeter such as LAr, the uncertainty associated with the light calorimetry calibration itself is expected to contribute as a constant term to the overall energy resolution budget~\cite{CalorimetryBibleFabjan}. The excellent energy resolution achieved from the dual calorimetry in this study also requires superb signal-to-noise ratio in both the charge and light readout systems of the LArTPC.

The excellent light collection capability from extensive coverage and nanosecond timing resolution will also reduce solar neutrino analysis thresholds in a LArTPC by requiring a coincidence of pulses expected in a golden channel in the daughter $^{40}\mathrm{K}^\ast$ deexcitation leading to very low background solar neutrino samples. The background rejection using coincident light signals should be tested with a robust model of photon simulation in a LArTPC. Furthermore, a better understanding of the scintillation reconstruction is needed, including vertex resolution from light topology and timing. Reconstruction of the delayed $\gamma$ pulse also requires a precise knowledge of the scintillation time profile (with or without Xe doping). The lifetime of the $^{40}$K state, $480$~ns, is short compared to the few-$\mu$s primary pulse. Therefore pulse-shape discrimination will be critical for identifying the two-pulse structure. How this affects energy and vertex resolution must also be studied.

The nonuniformity of light collection and its impact on energy resolution under different position reconstruction scenarios is summarized in Appendix~\ref{sec:nonuniformly}. This nonuniformity can be corrected once an event is located precisely by a combination of detected light signals and charge-light matching. As a result, the nonuniform LY after precise position reconstruction has little impact on energy smearing. 

This study provides guidance to the realization of an enhanced photon detection system in future LArTPCs with high and uniform LY. Furthermore, this work will motivate the development of more sophisticated techniques and analyses utilizing both the charge and the light detection systems in future LArTPCs to realize these low energy physics prospects.

\bigskip

\begin{acknowledgments}
We are grateful to have L. Strigari and Y. Zhuang from TAMU provide the latest estimated SURF atmospheric flux. We thank E. Church, E. Granados-Vazquez, C. Jung, X. Qian, J. Reichenbacher, K. Scholberg, and L. Wan for their insightful discussions on the studies presented in this paper. This work is supported by 2024 SBU-BNL Seed Grant Program and BNL LDRD 23-058.
\end{acknowledgments}

\appendix
\section{Energy Resolution for MeV $\nu_{e}$-Ar CC Events without Hadron Emission}
\label{sec:Eresnohadrons}

\begin{figure}[ht]
  \centering  \includegraphics[width=0.9\columnwidth]{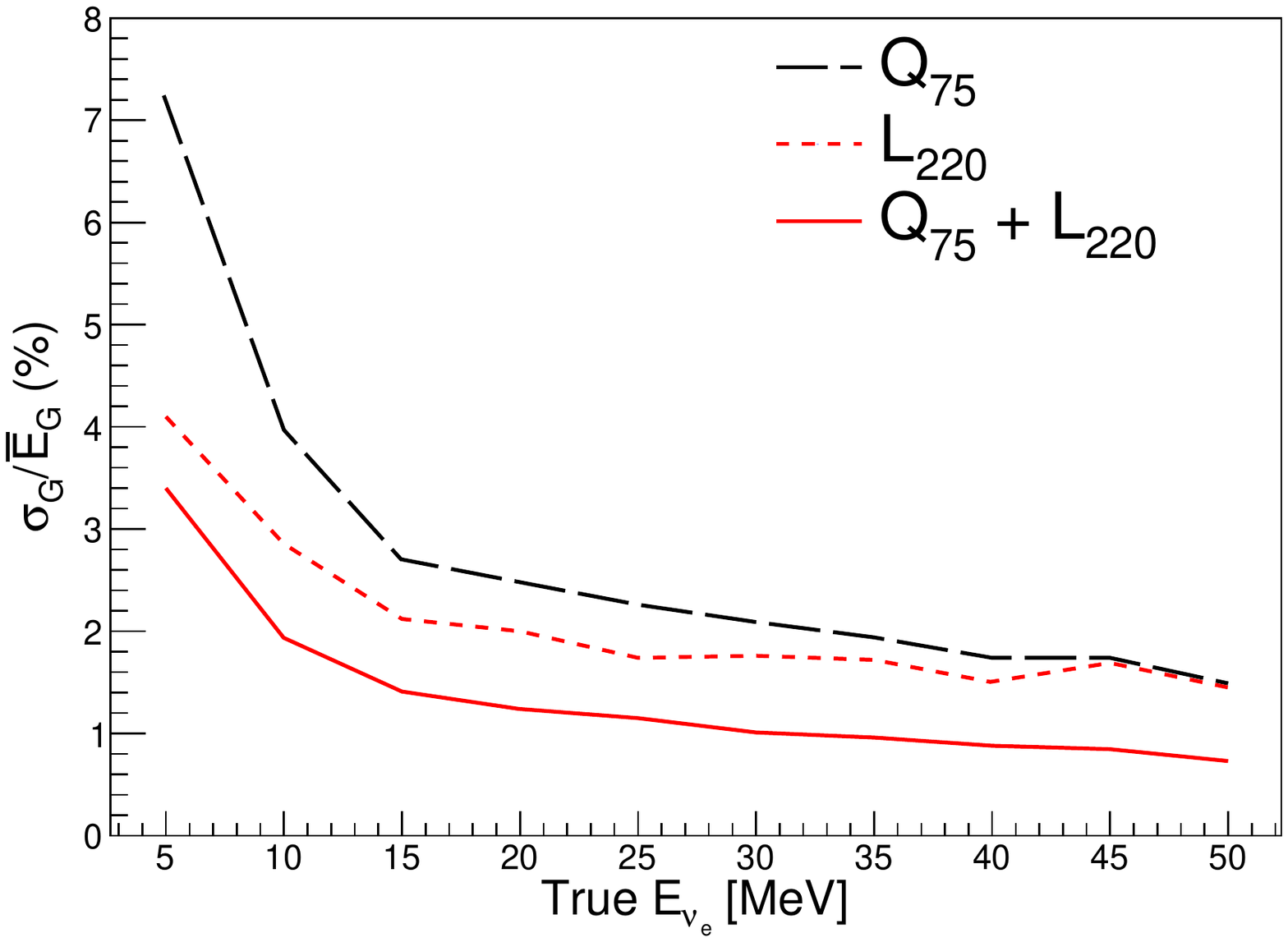}
  \centering  \includegraphics[width=0.9\columnwidth]{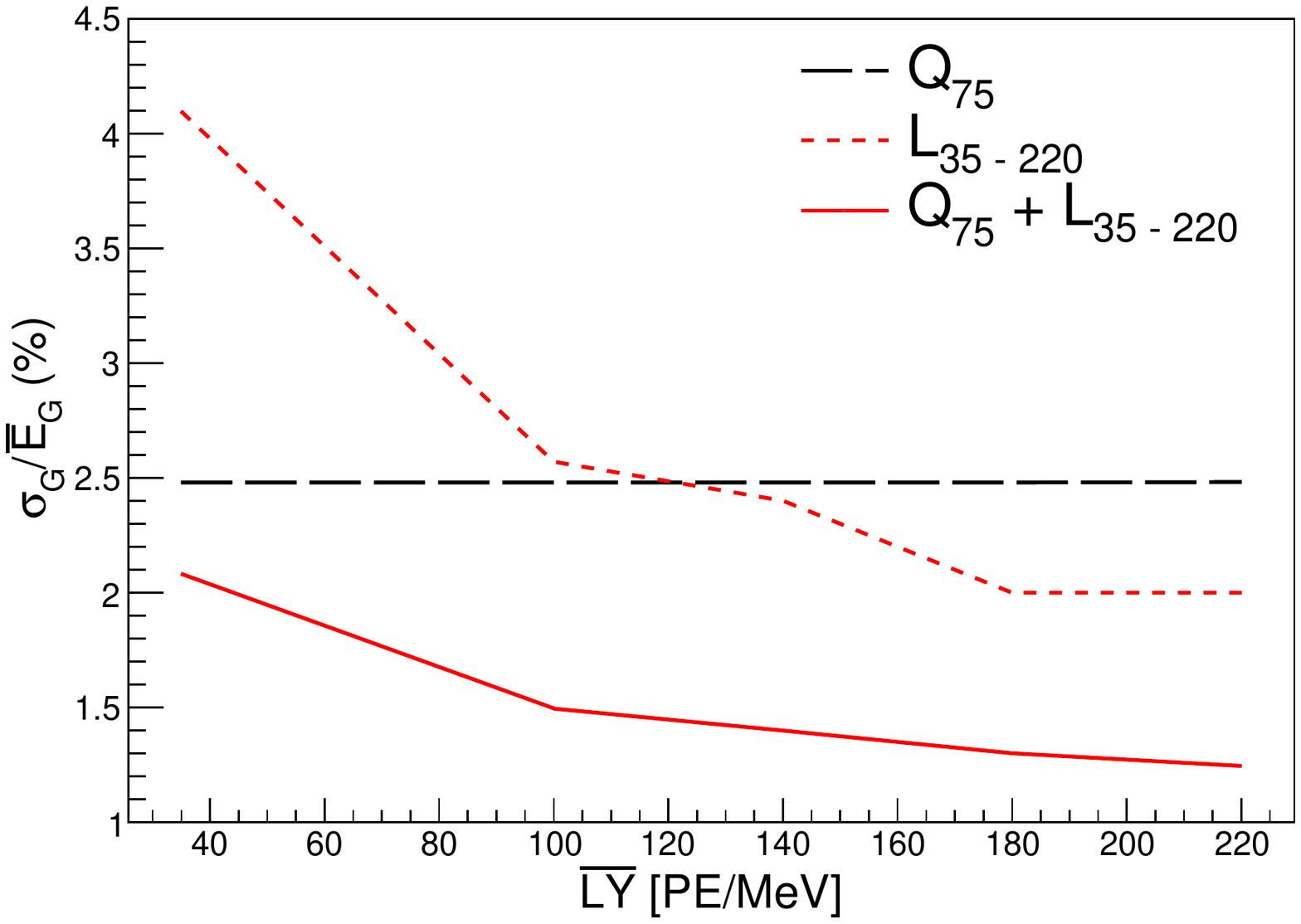}
  \caption{Reconstructed energy resolution for the primary peak containing only events without hadrons. Top: reconstructed ${E}_{\text{avail}}$ resolution as a function of true ${E}_{\nu_{e}}$ from 5 to 50 MeV. Bottom: reconstructed ${E}_{\text{avail}}$ resolution as a function of benchmark $\overline{\text{LY}}$ for 20 MeV $\nu_{e}$.}
  \label{fig:RecoEresFit}
\end{figure}

To highlight the improvement in energy resolution provided by combined calorimetry for neutrino events without hadrons, we fit a Gaussian function to the primary reconstructed energy peak, excluding events with hadron emissions. We use the fitted standard deviation ($\sigma_{G}$) and mean energy ($\bar{E}_{G}$) to quantify the energy resolution. The energy resolution as a function of true ${E}_{\nu_{e}}$ and as a function of $\overline{\text{LY}}$ for the 20 MeV $\nu_{e}$ sample is shown in Fig.~\ref{fig:RecoEresFit}, based on the Gaussian-fitted primary peak.

In Fig.~\ref{fig:RecoEresFit}, for the 35 MeV $\nu_{e}$-Ar CC sample, the combined calorimetry achieves an energy resolution 1.0\% (5.9\%/$\sqrt{{E}}$), compared to 1.9\% (11.3\%/$\sqrt{{E}}$) for charge-only calorimetry.  For the primary peak of events without hadrons, the energy resolution of the light-only reconstruction at $\overline{\text{LY}} = $ 220 PE/MeV already surpasses that of charge-only reconstruction across true ${E}_{\nu_{e}}$ range of 5-50 MeV. Although not shown here, this trend also holds for $\overline{\text{LY}}$ above 140 PE/MeV. The combined calorimetry further improves the energy resolution across the same energy range.

Here the energy resolution is impressively low because it only considers events without any hadron emission. All hadron tagging related assumptions in the main text such as the neutron capture don’t affect this result. In addition, we only considered statistical term in the light only energy resolution. The typical contributions related to light calorimetry calibration and dark noise are neglected. For the charge side, the resolution contribution primarily comes from the applied detection threshold (cf. Fig~\ref{fig:DetectedQL} top plot) and recombination. Effects such as charge cluster and reconstruction are neglected.

For the 20 MeV $\nu_{e}$ sample shown in the bottom plot of Fig.~\ref{fig:RecoEresFit}, the energy resolution of the light-only reconstruction ${E}_{\text{reco}, \ {L}_{220}}$ is 2.0\% (8.8\%/$\sqrt{{E}}$), while the resolution from charge-only calorimetry ${E}_{\text{reco}, \ {Q}_{75}}$ is 2.5\% (10.8\%/$\sqrt{{E}}$). When $\overline{\text{LY}}$ exceeds 100 PE/MeV, the energy resolution of light-only calorimetry (dotted lines) is comparable or better than that of charge-only calorimetry ${E}_{\text{reco}, \ {Q}_{75}}$ (dashed line). Furthermore, across all benchmark $\overline{\text{LY}}$ scenarios, combined charge and light calorimetry improves energy resolution by approximately a factor of two compared to charge-only or light-only reconstruction. This enhancement is attributed to the intrinsic anti-correlation between charge and light signals.

Most importantly, with a significantly higher $\overline{\text{LY}}$ above 100 PE/MeV the reconstructed energy resolution can be up to twice as good as that in the 35 PE/MeV scenario, whether using light-only reconstruction or combined charge and light reconstruction.

\section{Nonuniform LY Impacts}
\label{sec:nonuniformly}

LY is not uniform across the detector volume due to effects such as optical coverage, photon propagation, and detector efficiency variations. A Geant4 simulation of the light generation and propagation in the proposed APEX design (Fig.~\ref{fig:apexlymap}) provides the LY values at different positions in the detector with an average LY of 180 PE/MeV and minimum LY of 109 PE/MeV. If an event can be precisely located in this detector, the LY nonuniformity can be corrected based on the calibrated LY map. However, location uncertainties arise when using solely the photon detection system due to the photodetector granularity. This position uncertainty introduces a mismatch between the LY for event calorimetric reconstruction and the LY for the initial generation of photons. 

\begin{figure}[ht]
  \centering  \includegraphics[width=0.9\columnwidth]{APEX_LY.png}
  \caption{Map excerpted from Ref.~\cite{DUNE_Phase2} showing the expected LY in the central (x, y) transverse plane at z = 0 for an extended photon detection system instrumented on the field cage. Dimmer regions are present near the anode planes (with no photon detectors) and at mid-height (near the non-instrumented cathode plane).}
  \label{fig:apexlymap}
\end{figure}	

We compare three scenarios considering the treatment of LY and its nonuniformity during the initial photon signal generation and the final reconstruction of deposited energy in light. The first scenario is the basic assumption of a uniform LY of 180 PE/MeV as described in the main text. The second scenario considers position-dependent LY following the map in Fig.~\ref{fig:apexlymap} when generating photons and reconstructing the detected light. In this case, we assume the event position can be precisely reconstructed so that the detected energy in light can be simulated using the same LY that's used for the initial photon generation. The precise location of an MeV event is possible after a proper and efficient charge-light (Q-L) matching. In this scenario, Eq.~\ref{eq:NPE} for initial photon generation now becomes
\begin{equation}\label{eq:NPE_case2}
\overline{{N}}_{\text{ph}} = \frac{{dL}}{19.5 \ \text{eV}} \times \text{PCE(LY(x, y))}
\end{equation}
where LY(x, y) represents the LY used to generate photons for an energy deposit at a location (x, y). And Eq.~\ref{eq:detectedlight} for reconstructed energy deposit in light becomes
\begin{equation}\label{eq:detectedlight_case2}
{L}_{\text{LY}} = \frac{{N}_{\text{PE}} \times 19.5 \ \text{eV}}{\text{PCE}_{\text{Q-L\ matching}}}
\end{equation}
where $\text{PCE}_{\text{Q-L\ matching}}$ represents the photon collection efficiency assuming a precise event position can be reconstructed after Q-L matching and
\begin{equation} 
    \text{PCE}_{\text{Q-L\ matching}} = \text{PCE}(\text{LY(x, y)}).
    \label{pce_case2}
\end{equation}

\begin{figure}[t!]
  \centering  \includegraphics[width=0.9\columnwidth]{nonuniformLY_Lonly.png}
  \centering  \includegraphics[width=0.9\columnwidth]{nonuniformLY_QL.png}
  \centering  \includegraphics[width=0.9\columnwidth]{nonuniformLY_QL_withhadrontag.png}
  \caption{Reconstructed total available neutrino energy for true $E_\nu$ = 35 MeV with different treatment of LY and its nonuniformity: a uniform LY of 180 PE/MeV as in the main text (blue); precise position reconstruction with charge-light matching (red); light-only position reconstruction (green). Top: analysis B light-only calorimetry. Middle: analysis B combined calorimetry. Bottom: analysis C combined calorimetry.}
  \label{fig:nonuniformLY_Ereco}
\end{figure}

In the last scenario, we consider both the position-dependent LY and reconstructed position uncertainties using solely light detectors. This represents the worst case and we expect to see the impact from the nonuniform light collection. Here the LY for initial photon generation is the same as the second scenario in Eq.~\ref{eq:NPE_case2}. 
\begin{figure}[t!]
  \centering  \includegraphics[width=0.9\columnwidth]{nonuniformLY_Eres_GaussFit_Lonly.png}
  \centering  \includegraphics[width=0.9\columnwidth]{nonuniformLY_Eres_GaussFit_QL.png}
  \centering  \includegraphics[width=0.9\columnwidth]{nonuniformLY_Eres_GaussFit_QL_withhadrontag.png}
  \caption{Reconstructed energy resolution $\sigma_G/\bar{E}_G$ for events without hadron emission reported in Appendix~\ref{sec:Eresnohadrons} with different treatment of LY and its nonuniformity: a uniform LY of 180 PE/MeV as in the main text (blue); precise position reconstruction with charge-light matching (red); light-only position reconstruction (green). Top: analysis B light-only calorimetry. Middle: analysis B combined calorimetry. Bottom: analysis C combined calorimetry.}
  \label{fig:nonuniformLY_sigmean}
\end{figure}

\begin{figure}[t!]
  \centering  \includegraphics[width=0.9\columnwidth]{nonuniformLY_Eres_rmsmean_Lonly.png}
  \centering  \includegraphics[width=0.9\columnwidth]{nonuniformLY_Eres_rmsmean_QL.png}
  \centering  \includegraphics[width=0.9\columnwidth]{nonuniformLY_Eres_rmsmean_QL_withhadrontag.png}
  \caption{Reconstructed energy resolution $\sigma_h/\bar{E}_h$ for all events reported in the main text with different treatment of LY and its nonuniformity: a uniform LY of 180 PE/MeV as in the main text (blue); precise position reconstruction with charge-light matching (red); light-only position reconstruction (green). Top: analysis B light-only calorimetry. Middle: analysis B combined calorimetry. Bottom: analysis C combined calorimetry.}
  \label{fig:nonuniformLY_rmsmean}
\end{figure}
However, the LY used to calculate detected energy in light is:
\begin{equation}\label{eq:detectedlight_case3}
{L}_{\text{LY}} = \frac{{N}_{\text{PE}} \times 19.5 \ \text{eV}}{\text{PCE}_{\text{Light only reconstruction}}}
\end{equation}
where
\begin{equation} 
    \text{PCE}_{\text{Light only reconstruction}} = \frac{\text{LY}(x_{s}, y_{s})}{21622}.
    \label{newpce_withunc}
\end{equation}   
Here $\text{LY}(x_{s}, y_{s})$ represents a LY sampled at a reconstructed position approximated as a Gaussian smeared true event position: $x_s \sim N(x_{\text{true}},\ \sigma_x), \quad y_s \sim N(y_{\text{true}},\ \sigma_y)$, where resolution \(\sigma_x\) and \(\sigma_y\) are the reconstructed position resolution. We take $\sigma_{x} = 28$ cm and $\sigma_{y} = 47$ cm for 4 MeV point energy deposit source using solely the APEX photon detectors for position reconstruction from Ref.~\cite{MarinhoAPEXsim}. These uncertainties are conservative since we look at true $E_\nu$ over 5 MeV.

The reconstructed total available energy distributions under the three scenarios of LY treatment for 35 MeV $\nu$-Ar CC events are shown in Fig.~\ref{fig:nonuniformLY_Ereco}. The main text analysis B light-only calorimetry (Fig.~\ref{fig:nonuniformLY_Ereco} top), analysis B combined charge and light calorimetry (Fig.~\ref{fig:nonuniformLY_Ereco} middle), and analysis C combined charge and light calorimetry with hadron PID for binding energy corrections (Fig.~\ref{fig:nonuniformLY_Ereco} bottom) are plotted separately. The distribution with the light-only position reconstruction uncertainty (green curve) are much broader than the other two scenarios. A precise position reconstruction assumed possible via charge-light matching (red) removes the effect on position dependendt LY and result in a similar performance to the uniform LY scenario (blue).

Fig.~\ref{fig:nonuniformLY_sigmean} and Fig.~\ref{fig:nonuniformLY_rmsmean} present energy resolutions $\sigma_G/\bar{E}_G$ and $\sigma_h/\bar{E}_h$ defined in the main text for neutrino energy from 5 to 50 MeV. The resolution with nonuniform LY and the uncertainty of the position of the event vertex has a larger impact on light-only calorimetric reconstruction than the combined calorimetric reconstruction.

Lastly, we comment on the precise position reconstruction scenario using charge-light matching (red curves in all plots) where the effect LY nonuniformity is removed. With an enhanced light detection system such as APEX, the light-only reconstructed position resolution is on the order of tens of cm. We expect once the interesting activity is located to this tens-of-cm region solely by the light detector, we can perform more efficient and accurate charge-light matching to pin point the activity more precisely, on the order of 1 cm (i.e., LArTPC charge readout position resolution). This helps us to correct for the LY nonuniformity effects.

\section{Energy Smearing Matrices}
\label{sec:Esmearmatrices}
\begin{figure}[h!]
  \centering
    \includegraphics[width=0.48\columnwidth]{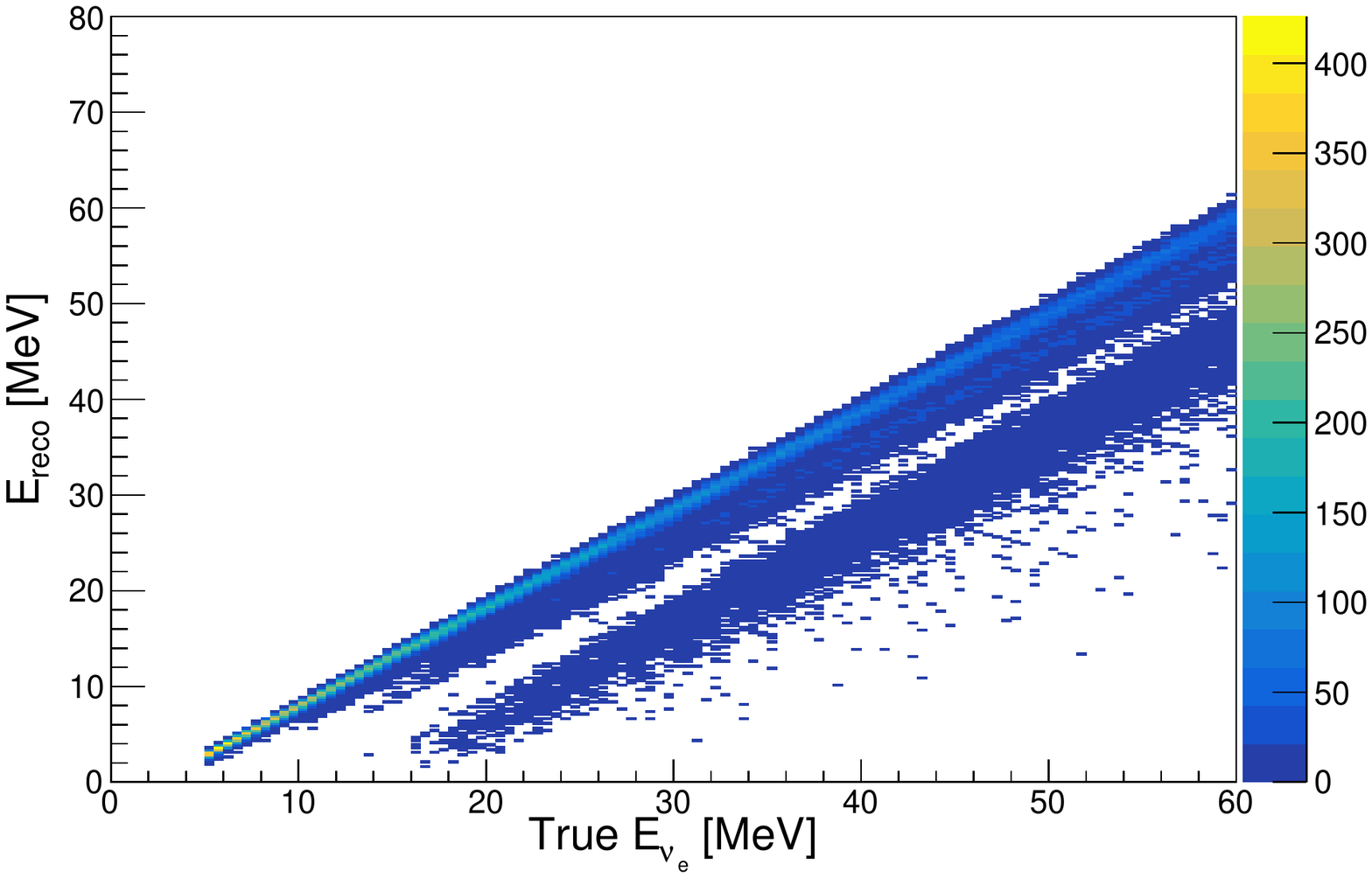}
    \includegraphics[width=0.48\columnwidth]{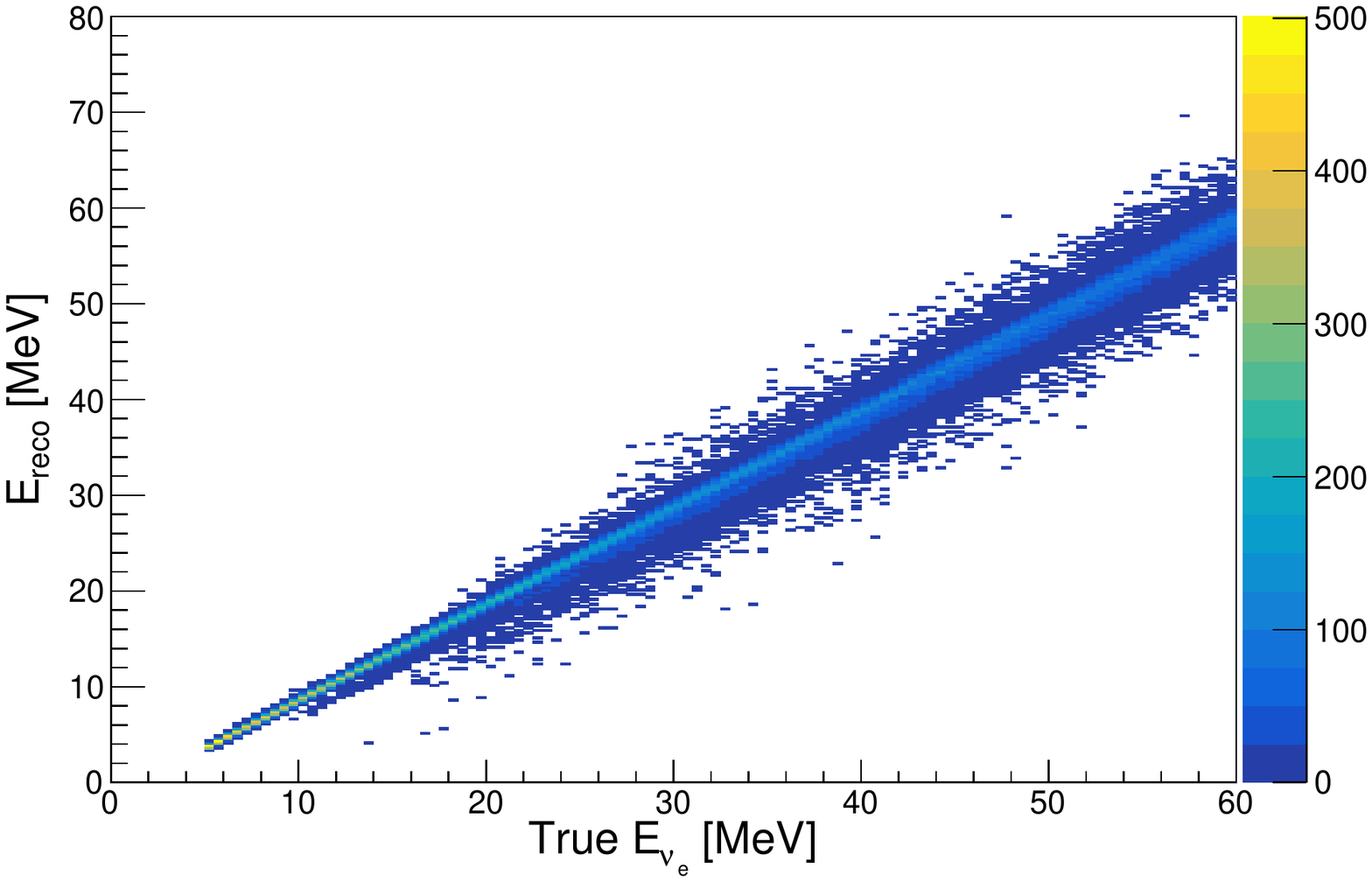}
    \includegraphics[width=0.48\columnwidth]{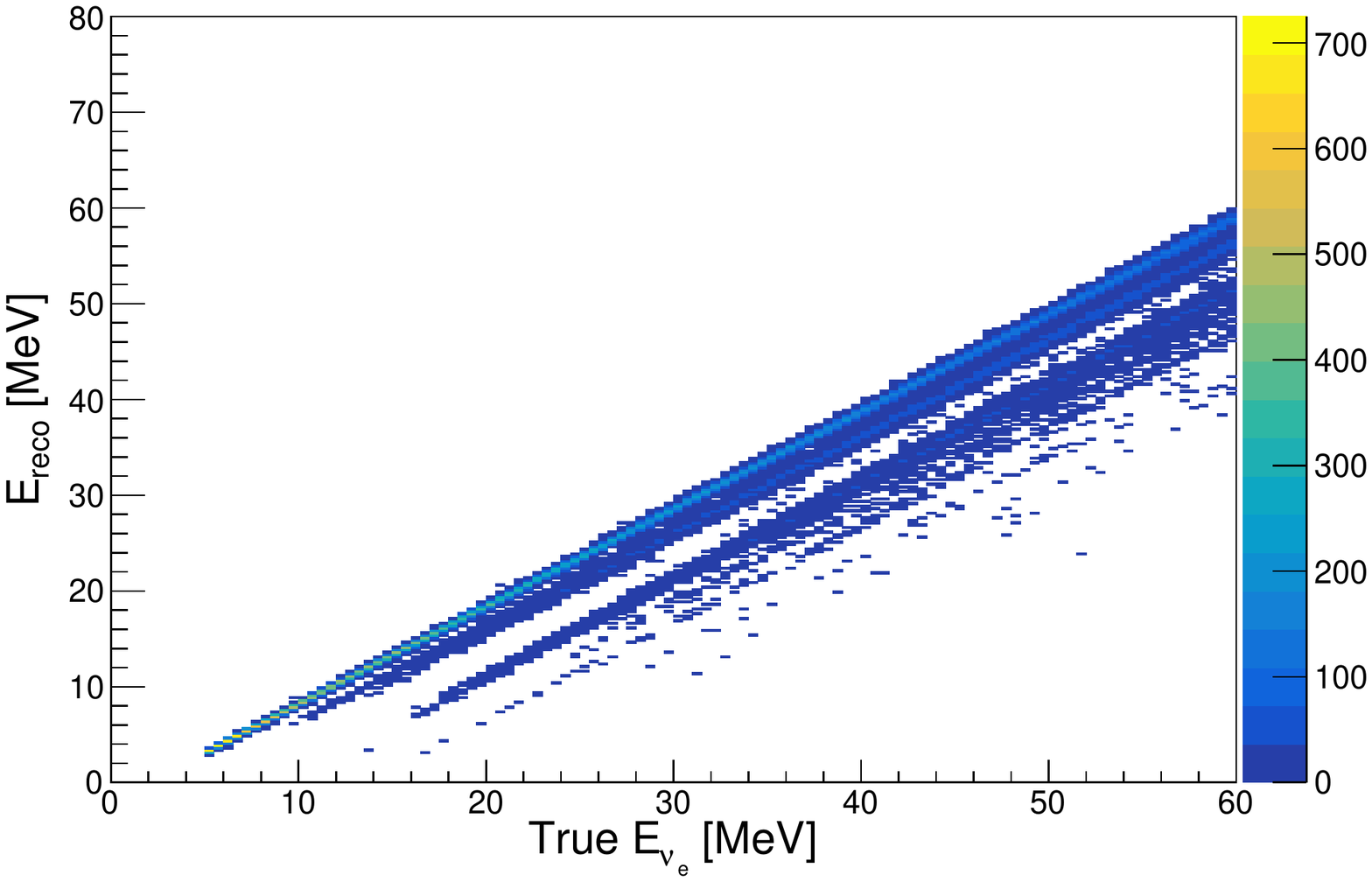}
    \includegraphics[width=0.48\columnwidth]{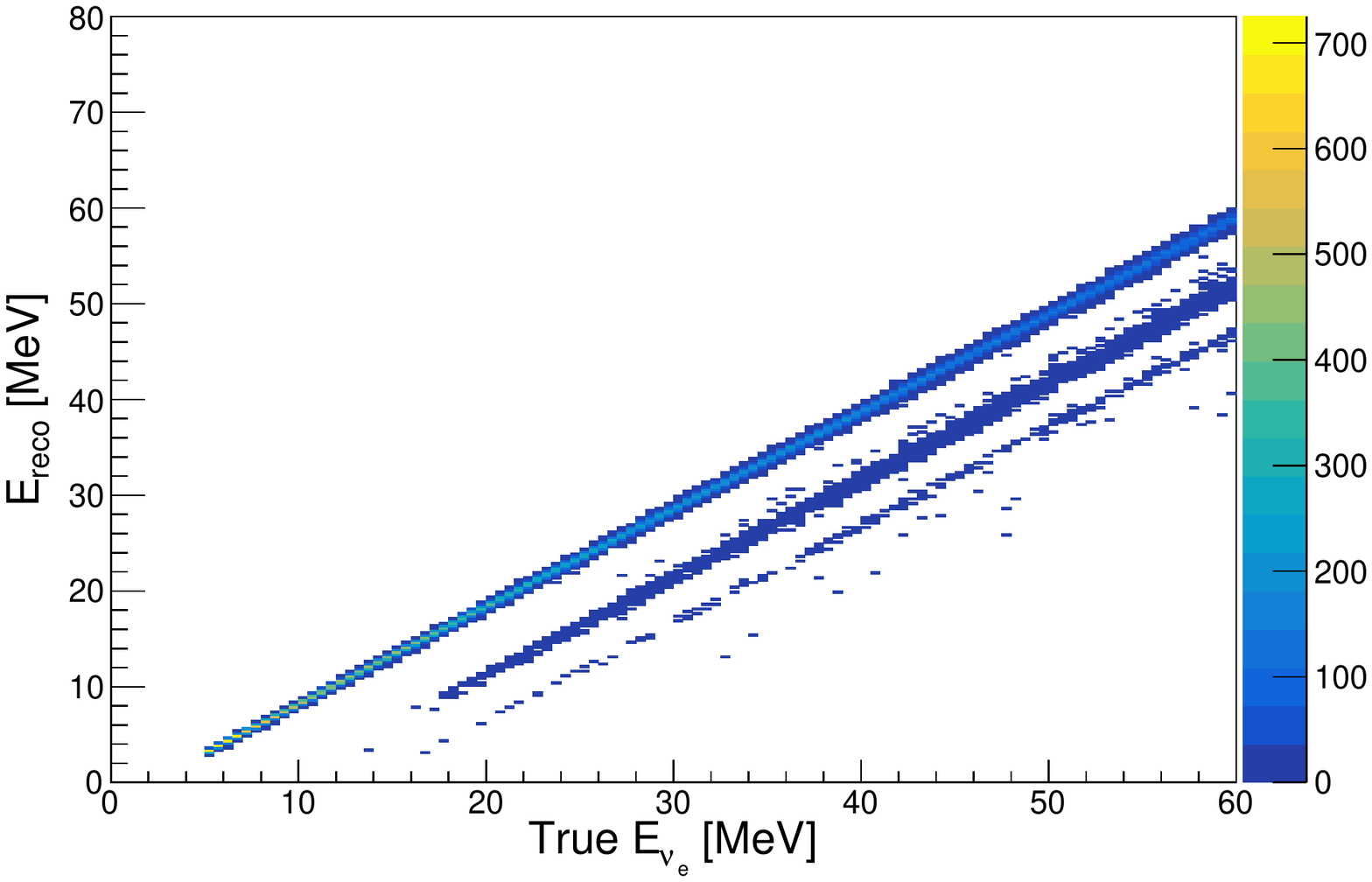}
  \caption{Energy response matrices of reconstructed neutrino energy vs.~true neutrino energy for electron neutrinos. The reconstructed energy comes from the charge-only deposited energy with 75 keV detection threshold, ${Q}_{75}$ (top left),  light-only ${L}_{180}$ (top right), combined charge and light deposited energy for $\overline{\text{LY}}$ = 180 PE/MeV, ${Q}_{75}\ +\ {L}_{180}$, without (bottom left), and with (bottom right) hadron tagged events removed.}
  \label{fig:response_matrix}
\end{figure}

Fig.~\ref{fig:response_matrix} shows 2D energy response matrices of reconstructed neutrino energy vs.~true neutrino energy for electron neutrinos. The reconstructed energy comes from the charge-only deposited energy with 75 keV detection threshold, ${Q}_{75}$ (top left),  light-only ${L}_{180}$ (top right), combined charge and light deposited energy for $\overline{\text{LY}}$ = 180 PE/MeV, ${Q}_{75}\ +\ {L}_{180}$, without (bottom left), and with (bottom right) hadron tagged events removed.

\bibliography{main}
\end{document}